\documentclass[11pt]{article}
\usepackage{amssymb,amsmath,amsfonts}
\usepackage{graphicx}
\usepackage{graphics}
\usepackage{eepic,epsfig}

\makeatletter \@addtoreset{equation}{section}

\makeatother

\begin{document}

\title{Bulk Casimir densities and vacuum interaction forces in higher dimensional brane models}

\author{Aram A. Saharian\thanks{%
Email: saharyan@server.physdep.r.am} \\
%EndAName
\textit{Department of Physics, Yerevan State
University, 1 Alex Manoogian Str.} \\
\textit{ 375049 Yerevan, Armenia,}\\
\textit{and}\\
\textit{The Abdus Salam International Centre for Theoretical
Physics} \\
\textit{ 34014 Trieste, Italy }}
\date{\today}
\maketitle

\begin{abstract}
Vacuum expectation value of the energy-momentum tensor and the
vacuum interaction forces are evaluated for a massive scalar field
with general curvature coupling parameter satisfying Robin boundary
conditions on two codimension one parallel branes embedded in
$(D+1)$-dimensional background spacetime $AdS_{D_1+1}\times \Sigma $
with a warped internal space $\Sigma $. The vacuum energy-momentum
tensor is presented as a sum of boundary-free, single brane induced,
and interference parts. The latter is finite everywhere including
the points on the branes and is exponentially small for large
interbrane distances. Unlike to the purely AdS bulk, the part
induced by a single brane, in addition to the distance from the
brane, depends also on the position of the brane in the bulk. The
asymptotic behavior of this part is investigated for the points near
the brane and for the position of the brane close to the AdS horizon
and AdS boundary. The contribution of Kaluza-Klein modes along
$\Sigma $ is discussed in various limiting cases. The vacuum forces
acting on the branes are presented as a sum of the self-action and
interaction terms. The first one contains well known surface
divergences and needs a further renormalization. The interaction
forces between the branes are finite for all nonzero interbrane
distances and are investigated as functions of the brane positions
and the length scale of the internal space. We show that there is a
region in the space of parameters in which these forces are
repulsive for small distances and attractive for large distances. As
an example the case $\Sigma =S^{D_2}$ is considered. An application
to the higher dimensional generalization of the Randall-Sundrum
brane model with arbitrary mass terms on the branes is discussed.
Taking the limit with infinite curvature radius for the AdS bulk,
from the general formulae we derive the results for two parallel
Robin plates on background of $R^{(D_1,1)}\times \Sigma $ spacetime.

\end{abstract}

\bigskip


\bigskip

\section{Introduction}

\label{sec:introd}

Recent proposals of large extra dimensions use the concept of brane
as a sub-manifold embedded in a higher dimensional spacetime, on
which the Standard Model particles are confined. Braneworlds
naturally appear in string/M-theory context and provide a novel
setting for discussing phenomenological and cosmological issues
related to extra dimensions. The model introduced by Randall and
Sundrum \cite{Rand99a} is particularly attractive. The corresponding
background solution consists of two parallel flat 3-branes, one with
positive tension and another with negative tension embedded in a
five dimensional AdS bulk. The fifth coordinate is compactified on
$S^1/Z_2$, and the branes are on the two fixed points. It is assumed
that all matter fields are confined on the branes and only the
gravity propagates freely in the five dimensional bulk. In this
model, the hierarchy problem is solved if the distance between the
branes is about 37 times the AdS radius and we live on the negative
tension brane. More recently, alternatives to confining particles on
the brane have been investigated and scenarios with additional bulk
fields have been considered.

From the point of view of embedding the Randall-Sundrum model into a
more fundamental theory, such as string/M-theory, one may expect
that a more complete version of this scenario must admit the
presence of additional extra dimensions compactified on an internal
manifold. From a phenomenological point of view, higher dimensional
theories with curved internal manifolds offer a richer geometrical
and topological structure. The consideration of more general
spacetimes may provide interesting extensions of the Randall-Sundrum
mechanism for the geometric origin of the hierarchy. Spacetimes with
more than one extra dimension can allow for solutions with more
appealing features, particularly in spacetimes where the curvature
of the internal space is nonzero. More extra dimensions also relax
the fine-tunings of the fundamental parameters. These models can
provide a framework in the context of which the stabilization of the
radion field naturally takes place. In addition, a richer
topological structure of the field configuration in transverse space
provides the possibility of more realistic spectrum of chiral
fermions localized on the brane. Several variants of the
Randall--Sundrum scenario involving cosmic strings and other global
defects of various codimensions have been investigated in higher
dimensions (see, for instance, \cite{Greg00}-\cite{Davo03} and
references therein). In particular, much work has been devoted to
warped geometries in six dimensions.

Motivated by the problems of the radion stabilization and the
generation of cosmological constant, the role of quantum effects in
braneworlds has attracted great deal of attention (see, for
instance, references given in \cite{Saha05I}). In this paper we
continue the investigation of the local quantum effects induced by
two codimension one parallel branes embedded in the background
spacetime $AdS_{D_1+1}\times \Sigma $ with a warped internal space
$\Sigma $ (for the investigation of the vacuum energy-momentum
tensor in the geometry with $AdS_{D+1}$ bulk see Refs.
\cite{Knap03,Saha04a}). The quantum effective potential and the
problem of moduli stabilization in the orbifolded version of this
model with zero mass parameters on the branes are discussed recently
in Ref. \cite{Flac03b}. In particular, it has been shown that one
loop-effects induced by bulk scalar fields generate a suitable
effective potential which can stabilize the hierarchy without fine
tuning. In the previous paper \cite{Saha05I} we have studied the
Wightman function and the vacuum expectation value of the field
square for a scalar field with an arbitrary curvature coupling
parameter obeying Robin boundary conditions on the branes. For an
arbitrary internal space $\Sigma $, the application of the
generalized Abel-Plana formula allowed us to extract form the vacuum
expectation values the part due to the bulk without branes and to
present the brane induced parts in terms of exponentially convergent
integrals for the points away the branes. An application of the
general results to the model with $\Sigma =S^{N}$ was discussed.
Here we consider the vacuum expectation values of the
energy-momentum tensor and the vacuum interaction forces between the
branes. In addition to describing the physical structure of the
quantum field at a given point, the energy-momentum tensor acts as
the source in the Einstein equations and therefore plays an
important role in modelling a self-consistent dynamics involving the
gravitational field.

The paper is organized as follows. In the next section, by using the
expression for the Wightman function from \cite{Saha05I}, we derive
formula for the expectation values of the energy-momentum tensor for
a single brane geometry in both regions on the left and on the right
of brane. Various limiting cases are discussed in the general case
of the internal manifold. The corresponding quantities for the
geometry of two branes are investigated in section \ref{sec:EMT2pl}.
The  case of different Robin coefficients on separate branes is
considered and an application to the higher dimensional version of
the Randall-Sundrum braneworld is discussed. In section
\ref{sec:Intforce} we study the vacuum interaction forces between
the branes as functions on brane positions and AdS radius. The
application of the general formulae to the example with the internal
space $\Sigma =S^{1}$ with the further generalization to the case
$S^{D_2}$ is discussed in section \ref{sec:exampS1}. Section
\ref{sec:Conc} contains a summary of the work and some suggestions
for further research.

\section{Casimir densities for a single brane}

\label{sec:EMT1pl}

In this paper we are intereseted in local quantum effects arising
from a scalar field $\varphi (x)$ propagating on background of
$(D+1)$-dimensional spacetime with topology $AdS_{D_1+1}\times
\Sigma $ and the line element (we adopt the conventions of Ref.
\cite{Birrell} for the metric signature and the curvature tensor)
\begin{equation}
ds^{2}=g_{MN}dx^{M}dx^{N}=e^{-2k_Dy}\eta _{\mu \sigma }dx^{\mu
}dx^{\sigma }-e^{-2k_Dy}\gamma _{ik}dX^{i}dX^{k}-dy^{2},
\label{metric}
\end{equation}%
where $\eta _{\mu \sigma }={\rm diag}(1,-1,\ldots ,-1)$ is the
metric tensor for $D_{1}$-dimensional Minkowski spacetime
$R^{(D_1-1,1)}$, $k_D$ is the inverse AdS radius, and the
coordinates $X^{i}$, $i=1,\ldots ,D_{2}$, cover the manifold $\Sigma
$, $ D=D_{1}+D_{2}$. Here and below the upper-case latin indices
$M,N$ range from $0$ to $D$ and $\mu ,\sigma =0,1,\ldots ,D_{1}-1$.
In addition to the radial coordinate $y$ we will also use the
coordinate $z=e^{k_Dy}/k_D$, in terms of which line element
(\ref{metric}) is written in the form conformally related to the
metric in the direct product spacetime $R^{(D_1,1)}\times \Sigma $
by the conformal factor $(k_Dz)^{-2}$. The solutions to Einstein
equations of type (\ref{metric}) have been considered in Refs.
\cite{Gher00b,Rand00,Oda01}. For the scalar field with curvature
coupling parameter $\zeta $ the field equation has the form
\begin{equation}
\left( \nabla ^{M}\nabla _{M}+m^{2}+\zeta R\right) \varphi (x)=0,
\label{fieldeq}
\end{equation}%
with $R$ being the scalar curvature for the background spacetime and
$\nabla _{M}$ is the covariant derivative operator associated with
the metric tensor $g_{MN}$. For minimally and conformally coupled
scalars one has $\zeta =0$ and $\zeta =\zeta _{D}\equiv (D-1)/(4D)$,
respectively. In the case of the bulk under consideration, the
corresponding Ricci scalar is given by formula
$R=-D(D+1)k_{D}^2-e^{2k_Dy}R_{(\gamma )}$, where $R_{(\gamma )}$ is
the scalar curvature for the metric tensor $\gamma _{ik}$. In the
discussion below we will assume that the field obeys Robin boundary
conditions
\begin{equation}  \label{boundcond}
\left( \tilde A_y+\tilde B_y\partial _y\right) \varphi (x)=0,\quad
y=a,b,
\end{equation}
with constant coefficients $\tilde A_y$, $\tilde B_y$, on two
parallel branes of codimension one, located at $y=a$ and $y=b$,
$a<b$. The expressions for these coefficients in a higher
dimensional generalization of the Randall-Sundrum brane model will
be given below. The $z$-coordinates of the branes we denote by
$z_j=e^{k_Dj}/k_D$, $j=a,b$. Robin type conditions are an extension
of Dirichlet and Neumann boundary conditions and appear in a variety
of situations, including the considerations of vacuum effects for a
confined charged scalar field in external fields, spinor and gauge
field theories, quantum gravity and supergravity. These boundary
conditions naturally arise for scalar and fermion bulk fields in
braneworld models. The Wightman function and the vacuum expectation
value (VEV) of the field square for the geometry of two branes in
the bulk with line element (\ref{metric}) are investigated in our
previous paper \cite{Saha05I}. Here we are interested in the VEV of
the energy-momentum tensor and the vacuum interaction forces between
the branes.

The VEV of the energy-momentum tensor can be evaluated by
substituting the Wightman function and the VEV of the field square
into the formula
\begin{eqnarray}
\langle 0|T_{MN}(x)|0\rangle &=&\lim_{x^{\prime }\rightarrow
x}\partial _{M}\partial _{N}^{\prime }\langle 0|\varphi (x)\varphi
(x^{\prime })|0\rangle \nonumber \\
&& +\left[ \left( \zeta -\frac{1}{4}\right) g_{MN}\nabla _{L}\nabla
^{L} -\zeta \nabla _{M}\nabla _{N}-\zeta R_{MN}\right] \langle
0|\varphi ^{2}(x)|0\rangle , \label{vevEMT1pl}
\end{eqnarray}
with the components of the Ricci tensor (here and below the
components of tensors are given in the coordinate system with the
radial coordinate $y$)
\begin{equation}\label{RMN}
R_{\mu \sigma }=-Dk_D^2g_{\mu \sigma },\quad R_{ik}=R_{(\gamma
)ik}-Dk_D^2g_{ik},\quad R_{DD}=Dk_D^2,
\end{equation}
where $R_{(\gamma )ik}$ is the Ricci tensor for the metric $\gamma
_{ik}$. This corresponds to the point-splitting regularization
technique for the VEVs. First let us consider the geometry of a
single brane located at $y=a$. In the region $y>a$ the corresponding
Wightman function is determined by the formula \cite{Saha05I}
\begin{eqnarray}
\langle \varphi (x)\varphi (x^{\prime })\rangle &=&\frac{%
k_{D}^{D-1}(zz^{\prime })^{D/2}}{2^{D_1}\pi ^{D_1-1}}\sum_{\beta
}\psi _{\beta
}(X)\psi _{\beta }^{\ast }(X^{\prime })\int d{\bf k}\,e^{i{\bf k}({\bf x}-%
{\bf x}^{\prime })}  \nonumber \\
&\times & \bigg\{ \int_{0}^{\infty }du\,u\frac{e^{i(t^{\prime
}-t)\sqrt{u^{2}+k_{\beta }^{2}}}}{\sqrt{u^{2}+k_{\beta }^{2}}}J_{\nu
}(u z)J_{\nu
}(u z^{\prime })-\frac{2}{\pi }\int_{k_{\beta }}^{\infty }du\, u\frac{\bar{I}%
_{\nu }^{(a)}(uz_{a})}{\bar{K}_{\nu }^{(a)}(uz_{a})} \nonumber \\
&\times & \frac{K_{\nu }(uz)K_{\nu }(uz^{\prime
})}{\sqrt{u^{2}-k_{\beta }^{2}}}
\cosh \!\left[ (t-t^{\prime })\sqrt{u^{2}-k_{\beta }^{2}}%
\right] \bigg\} .  \label{WF1bran}
\end{eqnarray}%
with $k_{\beta }=\sqrt{k^2+\lambda _{\beta }^2}$, $k=|{\bf k}|$, and
${\bf k}$ is the wave vector in the subspace $R^{(D_1-1,1)}$ with
spatial coordinates ${\bf x}=(x^{1},\ldots ,x^{D_1-1})$. In formula
(\ref{WF1bran}), $J_{\nu }(x)$ is the Bessel function, $I_{\nu
}(x)$, $K_{\nu }(x)$ are the Bessel modified functions with the
order
\begin{equation}
\nu =\sqrt{D^2/4-D(D+1)\zeta +m^{2}/k_{D}^{2}},  \label{nu}
\end{equation}%
$\psi _{\beta }(X)$ are the eigenfunctions for the operator $\Delta
_{(\gamma )}+\zeta R_{(\gamma )}$ with eigenvalues $-\lambda _{\beta
}^2$, and $\Delta _{(\gamma )}$ is the Laplace-Beltrami operator for
the metric $\gamma _{ik}$. Here and below for a given function
$F(x)$ we use the notation
\begin{equation}
\bar{F}^{(j)}(x)=A_{j}F(x)+B_{j}xF^{\prime }(x), \label{notbar}
\end{equation}%
with the coefficients related to the constants in the boundary
conditions by the formulae
\begin{equation}
A_{j}=\tilde{A}_{j}+ \tilde{B}_{j}k_{D}D/2,\quad
B_{j}=\tilde{B}_{j}k_{D}. \label{AjBj}
\end{equation}

In formula (\ref{WF1bran}), the part with the first integral in the
figure braces is the Wightman function for $AdS_{D_1+1}\times \Sigma
$ spacetime without branes, and the part with the second integral is
induced in the region $y>a$ by the presence of the brane. The
formula for the Whightman function in the region $y<a$ is obtained
from Eq. (\ref{WF1bran}) by the interchange of the Bessel modified
functions, $I_{\nu }\rightleftarrows K_{\nu }$. Substituting the
Wightman function and the expression for the VEV of the field square
from Ref. \cite{Saha05I} into formula (\ref{vevEMT1pl}), the VEV of
the energy-momentum tensor is presented in the form
\begin{equation}
\langle 0|T_M^N|0\rangle  = \langle T_M^N\rangle ^{(0)}+ \langle
T_M^N\rangle ^{(a)}, \label{EMT41pl}
\end{equation}
where
\begin{equation}\label{EMTAdS}
  \langle T_{M}^{N}\rangle ^{(0)}=\frac{%
k_{D}^{D+1}z^{D}}{(4\pi )^{\frac{D_1}{2}}} \Gamma \left(
1-\frac{D_1}{2}\right) \sum_{\beta } |\psi _{\beta
}(X)|^2\int_{0}^{\infty }du \, u(u^2+\lambda _{\beta
}^2)^{\frac{D_1}{2}-1} F_{\beta M}^{(-)N}[J_{\nu }(uz)],
\end{equation}
is the VEV for the energy-momentum tensor in the background without
branes, and the term
\begin{equation}\label{TMNbetadef}
\langle T_M^N\rangle ^{(a)}= \sum_{\beta }|\psi _{\beta }(X)|^2
\langle T_M^N\rangle ^{(a)}_{\beta }
\end{equation}
is induced by a single brane at $z=z_a$, with the contribution of
the given KK mode $\beta $ along $\Sigma $ determined by the formula
\begin{equation}
\langle T_M^N\rangle ^{(a)}_{\beta }=- \frac{2k_D^{D+1}z^{D}}{(4\pi
) ^{\frac{D_1}{2}} \Gamma \left( \frac{D_1}{2}\right) } \int
_{\lambda _{\beta }}^{\infty } d u\, u
 (u^2-\lambda _{\beta }^2)^{\frac{D_1}{2}-1} \frac{\bar{I}_{\nu }^{(a)}(uz_a)}{\bar{K} _{\nu
}^{(a)}(uz_a)}F_{\beta M}^{(+)N}[K_{\nu }(uz)]. \label{EMT1bounda}
\end{equation}
To derive this formula we have used the relation
\begin{equation}\label{relintk}
\int_{0}^{\infty }dk\, \int_{k_{\beta }}^{\infty }du
\frac{uk^{n-1}f(u)}{(u^2-k_{\beta }^2)^{s}}=\frac{\Gamma \left(
\frac{n}{2}\right) \Gamma (1-s)}{2\Gamma \left(
\frac{n}{2}+1-s\right)}\int_{\lambda _{\beta }}^{\infty }du\,
\frac{uf(u)}{(u^2-\lambda _{\beta }^2)^{s-\frac{n}{2}}} ,
\end{equation}
with $s=-1/2,1/2$ and $n=D_1-1$. For a given function $g(v)$, the
functions $F_{\beta M}^{(\pm )N}[g(v)]$ in formulae (\ref{EMTAdS})
and (\ref{EMT1bounda}) are defined by the relations
\begin{eqnarray}\label{Fmu}
 F_{\beta \mu }^{(\pm )\sigma }[g(v)] &=& \delta _{\mu }^{\sigma }\left( \frac{1}{4}-\zeta
  \right) \left\{ z^2g^2(v)\eta _{\beta }(X)
   +2 v\frac{\partial }{\partial v}F[g(v)] +
   \frac{\pm v^2-z^2\lambda _{\beta }^2}{D_1(\zeta -1/4)} g^2(v)
   \right\} ,\\
F_{\beta D }^{(\pm )D }[g(v)] &=& \left( \frac{1}{4}-\zeta
  \right) z^2g^2(v)\eta _{\beta }(X)+ \frac{1}{2}\bigg[ - v^2g^{\prime 2}(v)
   \nonumber  \\
&&   +D(4\zeta -1) v g(v)g'(v)+
   \left( \frac{2m^2}{k_D^2}-\nu ^2\pm v^2\right) g^2(v)\bigg]
    , \label{FD}
\end{eqnarray}
for the components in the AdS part, and by the relations
\begin{eqnarray}
F_{\beta D }^{(\pm )i }[g(v)] &=& \frac{k_D}{2}z^2(1-4\zeta )F[g(v)]
\eta ^{i}_{\beta }(X) , \label{FiD}\\
F_{\beta i }^{(\pm )k }[g(v)] &=& z^2g^2(v) \frac{t _{\beta i
}^{k}(X)}{|\psi _{\beta }(X)|^2} +\frac{1}{2}\delta _{i }^{k
}(1-4\zeta )v\frac{\partial }{\partial v}F[g(v)], \label{Fi}
\end{eqnarray}
with $t _{\beta i }^{k}(X)=-\gamma ^{k l}t _{\beta i l}(X)$, for the
components having indices in the internal space. In these
expressions we use the following notations
\begin{eqnarray}
F[g(v)]&=& v g(v) g'(v)+\frac{1}{2}\left( D+\frac{4\zeta }{4\zeta
    -1}\right) g^2(v), \label{Fgv} \\
    \eta _{\beta }(X)&=&\frac{\triangle _{(\gamma )}|\psi _{\beta }(X)|^2}{|\psi _{\beta
   }(X)|^2} ,\quad \eta ^{i}_{\beta }(X)=-\gamma ^{ik}\frac{\partial _{k}|\psi _{\beta
   }(X)|^2}{|\psi _{\beta }(X)|^2}
    , \label{etaX} \\
    t _{\beta ik}(X)&=&\nabla _{(\gamma )i}\psi _{\beta }(X)\nabla _{(\gamma )k}\psi ^{\ast }_{\beta
    }(X) +\nonumber \\
    && \left[ \left( \zeta -\frac{1}{4}\right) \gamma _{ik} \triangle _{(\gamma )}-\zeta
    \nabla _{(\gamma )i}\nabla _{(\gamma )k}-\zeta R_{(\gamma )ik}\right]
    |\psi _{\beta }(X)|^2,\label{tbetik}
\end{eqnarray}
where $\nabla _{(\gamma )i}$ is the covariant derivative operator
associated with the metric tensor $\gamma _{ik}$. By using the
equation for the Bessel modified functions, the derivative in
expressions (\ref{Fmu}) and (\ref{Fi}) can also be presented in the
form
\begin{equation}\label{derF}
v\frac{\partial }{\partial v}F[g(v)]=v^2g^{\prime 2}(v)+\left(
D+\frac{4\zeta }{4\zeta -1}\right) vg(v)g'(v)+(v^2+\nu ^2)g^2(v) ,
\end{equation}
for $g(v)=I_{\nu }(v),K_{\nu }(v)$.

By a similar way, for the VEV induced by a single brane in the
region $z<z_a$ one obtains
\begin{equation}
\langle T_M^N\rangle ^{(a)}_{\beta }=- \frac{2k_D^{D+1}z^{D}}{(4\pi
) ^{\frac{D_1}{2}} \Gamma \left( \frac{D_1}{2}\right) }\int
_{\lambda _{\beta }}^{\infty } d u\, u
 (u^2-\lambda _{\beta }^2)^{\frac{D_1}{2}-1} \frac{\bar{K}_{\nu }^{(a)}(uz_a)}{\bar{I} _{\nu
}^{(a)}(uz_a)}F_{\beta M}^{(+)N}[I_{\nu }(uz)].
\label{Tik1plnewleft}
\end{equation}
In the case of the purely AdS bulk without an internal space, from
formulae (\ref{EMT1bounda}) and (\ref{Tik1plnewleft}) we obtain the
results derived in Ref. \cite{Saha04a}. As we see from the formulae
given above, for the general case of the internal subspace $\Sigma$,
the vacuum energy-momentum tensor is non-diagonal. In the case of
homogeneous internal space the contributions into the
energy-momentum tensor coming from the terms containing $\eta
_{\beta }(X)$ and $\eta _{\beta }^{i}(X)$ vanish. In particular, in
this case one has $\langle T_D^i\rangle ^{(a)}=0$. Note that, unlike
to the $AdS_{D+1}$ bulk, VEVs (\ref{EMT1bounda}) and
(\ref{Tik1plnewleft}) in addition to the ratio $z/z_a$ depend also
on the absolute position of the brane. For a one-parameter internal
space of linear size $L$ one has $\lambda _{\beta }\sim 1/L$. In
this case the brane induced part in the vacuum energy-momentum
tensor is a function on the ratios $L/z_a$ and $z/z_a$. The latter
is related to the proper distance of the observation point from the
brane by the equation
\begin{equation}
z/z_a=e^{k_D(y-a)}. \label{propdisz}
\end{equation}
From the point of view of an observer residing on the brane, the
physical size of the subspace $\Sigma$ is $L_a=Le^{-k_Da}$ and the
corresponding KK masses are rescaled by the warp factor, $\lambda
_{\beta }^{(a)}=\lambda _{\beta }e^{k_Da}$. Now we see that the
brane induced part in the vacuum energy-momentum tensor is a
function of the proper distance from the brane and  on the ratio
$L_a/(1/k_D)$ of the physical size of the internal space $\Sigma $
for an observer living on the brane to the AdS curvature radius. For
a fixed value of the distance of the observation point from the
brane, the scaling of $L$ is equivalent to the shift of the brane
position.

Note that if we consider a quantum scalar field with mass $m$ on
background of spacetime $R\times \Sigma $ with the line element
$ds^2=dt^2-\gamma _{ik}dx^idx^k$, when the VEVs of the spatial
components of the energy-momentum tensor are expressed through the
tensor (\ref{tbetik}) by the formula
\begin{equation}\label{0Sigma}
\langle 0_{\Sigma }|T_{ik}|0_{\Sigma }\rangle
=\frac{1}{2}\sum_{\beta } \frac{t_{\beta ik}(X)}{\sqrt{\lambda
_{\beta }^2+m^2}} ,
\end{equation}
where $|0_{\Sigma }\rangle $ is the amplitude for the corresponding
vacuum state. The expression for the brane-free part (\ref{EMTAdS})
is divergent and needs regularization with further renormalization.
As it was discussed in Ref. \cite{Saha05I} for the case of the field
square, this can be done combining the zeta function technique for
the series over $\beta $ and dimensional regularization for the
integral over $u$. The series over $\beta $ is expressed in terms of
the local zeta function for the operator $\triangle _{(\gamma )
}+\zeta R_{(\gamma )}-m^2$ and its derivatives. In section
\ref{sec:exampS1}, to evaluate the brane-free part of the vacuum
energy-momentum tensor for the example with $\Sigma =S^{1}$, we will
use the Abel-Plana formula which allows to extract the part
corresponding to the bulk $AdS_{D+1}$ and to present the remained
finite part in terms of exponentially convergent integrals. The
brane-induced parts (\ref{EMT1bounda}) and (\ref{Tik1plnewleft}) are
finite for the points away the brane and, hence, the renormalization
procedure is needed for the boundary-free part only.

For the comparison with the corresponding results in the case of
bulk spacetime $AdS_{D_1+1}$ when the internal space is absent, it
is useful in addition to the VEVs (\ref{EMT1bounda}) and
(\ref{Tik1plnewleft}) to consider the VEV integrated over the
subspace~$\Sigma $:
\begin{equation}\label{TMNintegrated}
 \langle T_{M}^{N}\rangle ^{(a)}_{{\mathrm{integrated}}}=e^{-D_2k_Dy}\int_{\Sigma }
 d^{D_2}X\sqrt{\gamma }\,  \langle T_M^N\rangle ^{(a)} .
\end{equation}
Note that due to the warp factor in this formula the volume of the
extra space $\Sigma $ exponentially decreases as one moves toward
the AdS horizon. Comparing this integrated VEV with the
corresponding formula from Ref. \cite{Saha04a}, we see that for a
homogeneous subspace $\Sigma $ the contribution of the zero KK mode
($\lambda _{\beta }=0$) into the integrated ${}^{\sigma }_{\mu }$--
and ${}^{D}_{D}$--components of the brane induced vacuum
energy-momentum tensor differs from the corresponding formulae in
the $AdS_{D_1+1}$ bulk by the replacement $D\to D_1$ in expressions
(\ref{Fmu}), (\ref{FD}) and in the definition of the order for the
modified Bessel functions: for the latter case $\nu \to \nu _1$ with
$\nu _1$ defined by Eq. (\ref{nu}) with $D$ replaced by $D_1$. Note
that for a scalar field with $\zeta \leq \zeta _{D+D_1+1}$ one has
$\nu \geq \nu _1$. In particular, this is the case for minimally and
conformally coupled scalars.

It can be checked that the both boundary-free and brane induced
parts in the VEV of the energy-momentum tensor obey the continuity
equation $\nabla _{N}T^N_{M}=0$, which for the geometry under
consideration takes the form
\begin{eqnarray}\label{conteq1}
  && z^{D+1}\frac{\partial }{\partial z}\left( z^{-D}T_D^D\right) +
  D_1 T_0^0+T_{i}^{i}+\frac{1}{k_{D}}\nabla _{(\gamma
  )i}T^{i}_{D}=0, \\
  && z^{D+1}\frac{\partial }{\partial z}\left( z^{-D}T_i^D\right) +
  \frac{1}{k_{D}}\nabla _{(\gamma  )k}T^{k}_{i}=0, \label{conteq2}
\end{eqnarray}
with $\nabla _{(\gamma )i}T^{i}_{D}=\partial _{i}(\sqrt{\gamma }
T^{i}_{D})/\sqrt{\gamma }$. In particular, for a homogeneous
internal space the second equation is satisfied trivially and the
last term on the left of the first equation vanishes. Note that we
also have the relation $\nabla _{(\gamma )k}t_{\beta i}^{k}=0$. By
using the equation for the eigenfunctions $\psi _{\beta }(X)$, it
can be seen that the following trace relation takes place
\begin{equation}\label{trintspace}
\frac{t_{\beta i}^{i}(X)}{|\psi _{\beta }(X)|^{2}}=-(D_2-1)\zeta
_{D_2-1}\eta _{\beta }(X)-\lambda _{\beta }^2.
\end{equation}
On the base of this relation we easily verify that for a conformally
coupled massless scalar the brane induced VEVs (\ref{EMT1bounda})
and (\ref{Tik1plnewleft}) are traceless. The trace anomalies are
contained in the boundary-free part only.

To clarify the dependence of the boundary induced part in the VEV of
the energy-momentum tensor for general case of the internal space,
it is useful to consider various limiting cases when the
corresponding formulae are simplified. In the discussion below for
these cases it is convenient to introduce the following functions
\begin{subequations}\label{FbetMNuv}
\begin{eqnarray}
 F_{\beta D}^{(l)D}(u,v)&=& \left( \zeta -\frac{1}{4}\right)
 \eta _{\beta }(X) -u^2\delta _{0}^{l} \label{FbetMNuv1}\\
 F_{\beta \mu }^{(l)\sigma }(u,v)&=& \delta _{\mu }^{\sigma }\left[ F_{\beta D}^{(1)D}(u,v)
 +\frac{u^2-v^2}{D_1}+(4\zeta -1)u^2\delta _{1}^{l} \right] \label{FbetMNuv2}\\
 F_{\beta D}^{(l)i}(u,v)&=& k_D \left[ D(\zeta -\zeta _{D})\delta _{0}^{l}-2n^{(a)}
 z u \delta _{1}^{l}\left( \zeta -\frac{1}{4}\right)
 \right] \eta _{\beta }^{i}(X) , \label{FbetMNuv3} \\
F_{\beta i}^{(l)k}(u,v)&=& (4\zeta -1)u^2\delta _{i}^{k}\delta
_{1}^{l}-\frac{t _{\beta i }^{k}(X)}{|\psi _{\beta }(X)|^2},
\label{FbetMNuv4}
\end{eqnarray}
\end{subequations}
with $l=0,1$. Here and below we define $n^{(j)}=1$ for the region
$y>j$ and $n^{(j)}=-1$ for the region $y<j$. First of all, as a
partial check of the results derived above for the VEV of the
energy-momentum tensor let us consider the limit $k_D\to 0$. This
corresponds to a single Robin plate on the bulk $R^{(D_1,1)}\times
\Sigma $. For $k_D\to 0$, from (\ref{nu}) we see that the order $\nu
$ of the cylindrical functions in formulae (\ref{EMT1bounda}) and
(\ref{Tik1plnewleft}) is large. Introducing the new integration
variable $v=u/\nu $, we can use the uniform asymptotic expansions of
the Bessel modified functions for large values of the order (see,
for instance, \cite{abramowiz}). To the leading order this gives:
\begin{eqnarray}\label{TMNsplMink}
    \langle T_M^N\rangle ^{(a)}& \approx & \langle T_M^N\rangle
    ^{(a)}_{R^{(D_1,1)}\times \Sigma }=\frac{(4\pi )^{-
    \frac{D_1}{2}}}{\Gamma \left( \frac{D_1}{2}\right)}\sum_{\beta
}\left| \psi _{\beta }(X)\right| ^2 \nonumber \\
&& \times \int_{v_{\beta }}^{\infty } du\, (u^2-v_{\beta
}^{2})^{\frac{D_1}{2}-1}\frac{e^{-2u|y-a|}}{\tilde c_{a}(u)}
F_{\beta M}^{(1)N}(u,v_{\beta }),
\end{eqnarray}
where $v_{\beta }=\sqrt{m^2+\lambda _{\beta }^2}$ and in the
expression for $F_{\beta D}^{(1)i}(u,z)$ we should take $k_Dz=1$ (as
we are considering the limit $k_D\to 0$). In Eq. (\ref{TMNsplMink})
we have introduced the notation
\begin{equation}\label{cj}
    \tilde c_{j}(u)=\frac{\tilde A_j-n^{(j)}\tilde B_j u}{\tilde A_j+n^{(j)}\tilde B_j
    u}, \quad j=a,b.
\end{equation}
In particular, for a homogeneous internal space, from
(\ref{TMNsplMink}) we see that ${}^{D}_{D}$--component of the brane
induced part in the VEV of the energy-momentum tensor vanishes. For
Dirichlet (Neumann) scalar with $\zeta \leq \zeta _{D_1}$ the
corresponding energy density is negative (positive). In the special
case of bulk when the internal space $\Sigma $ is absent, formula
(\ref{TMNsplMink}) for the VEV $\langle T_M^N\rangle
^{(a)}_{R^{(D_1,1)}\times \Sigma }$ coincides with the result
previously derived in Ref. \cite{Rome02}.

For large values $u$, the integrands in formulae (\ref{EMT1bounda})
and (\ref{Tik1plnewleft}) behave as $u^{D_1}e^{-2u|z-z_a|}$ and the
integrals converge for the points away the branes. For the points on
the branes the integrals are divergent, leading to the divergent
VEVs of the energy-momentum tensor. These divergences are well-known
in quantum field theory with boundaries and are investigated in
general case of boundary geometry. To remove them more realistic
model for the brane is needed (see, for instance, the discussion in
Ref. \cite{Saha05I} and a model with the finite thickness brane in
Ref. \cite{Mina05}). To find the asymptotic behavior of the vacuum
energy-momentum tensor we note that near the brane the main
contribution into the $u$-integral comes from large values $u$, and
the Bessel modified functions can be replaced by their asymptotic
expressions for large values of the argument \cite{abramowiz}.
Assuming $k_D|y-a|\ll 1$ and $\lambda _{\beta }|z-z_a|\ll 1$ (note
that the second condition can also be written in the form $\lambda
_{\beta }^{(a)}|y-a|\ll 1$, the distance from the brane is much less
than the physical length scale of the KK mode), to the leading order
for the contribution of a given KK mode along $\Sigma $ one finds
the following result
\begin{equation}\label{TMNsplnear}
  \langle T_M^N\rangle ^{(a)}_{\beta }\approx \Gamma
  \left( \frac{D_1+1}{2}\right) \frac{\kappa (B_a)(k_Dz_a)^{D+1}F_{\beta M}^{N}}{2^{D_1}
  \pi ^{\frac{D_1+1}{2}}|z-z_a|^{D_1+1}},
\end{equation}
where $\kappa (B_j)=2\delta _{0B_j}-1$ and the coefficients for
separate components are defined by the formulae
\begin{subequations} \label{FbetaMN}
\begin{eqnarray}
&& F_{\beta \mu }^{\sigma } = D_1\delta _{\mu }^{\sigma } (\zeta
-\zeta _{D_1}), \quad F_{\beta D}^{D} =D(\zeta -\zeta _{D})
   \left( \frac{z}{z_a}-1\right) , \label{FbetaMN1near}\\
 & & F_{\beta D}^{i} =\left( \frac{1}{4}-\zeta \right) k_Dz_a
 (z-z_a) \eta _{\beta }^{i} (X) , \quad
   F_{\beta i}^{k} = D_1\delta _{i}^{k} \left(\zeta -\frac{1}{4}\right)
   . \label{FbetaMN2near}
\end{eqnarray}
\end{subequations}
As the renormalized boundary-free energy-momentum tensor is finite
on the brane, we conclude that near the brane the total vacuum
energy-momentum tensor is dominated by the brane induced part and
has opposite signs for Dirichlet and non-Dirichlet boundary
conditions. Near the brane ${}^{D}_{D}$-- and
${}^{i}_{D}$--components of this tensor have opposite signs in the
regions $y<a$ and $y>a$. For $\zeta
>\zeta _{D_1}$ ($\zeta <\zeta _{D_1}$) the energy density is
positive (negative) for Dirichlet boundary condition and is negative
(positive) for non-Dirichlet boundary condition. Note that for a
scalar field conformally coupled in $D$ spatial dimensions one has
$\zeta =\zeta _D>\zeta _{D_1}$. If we denote by $p_{i}$ and $p$ the
vacuum effective pressures in the subspace $\Sigma $ and in the
radial direction, respectively, when near the brane the equation of
state has the form $p_{i}=-D_1 \varepsilon /(D_1-1)$ for a minimally
coupled scalar and $p_{i}=D_1 \varepsilon /D_2$ for a conformally
coupled scalar, where $\varepsilon $ is the vacuum energy density.
In both cases one has $|p/\varepsilon |\ll 1$. Note that, unlike to
the case of the purely AdS bulk without an internal space, here the
vacuum energy-momentum tensor for a conformally coupled massless
scalar field diverges on the brane.

Next we consider the behavior of the brane induced VEV at large
distances from the brane. For nonzero KK modes along $\Sigma $
assuming $z\gg \lambda _{\beta }^{-1}$ we see that for the region of
integration in Eq. (\ref{EMT1bounda}) one has $uz\gg 1$ and the
MacDonald function can can be replaced  by its asymptotic expression
for large values of the argument. The main contribution into the
integral comes from values near the lower limit of integration and
to the leading order the brane induced part is estimated by the
formula
\begin{equation}\label{phi2spllargez}
\langle T_M^N\rangle ^{(a)}_{\beta }\approx
\frac{(k_Dz)^{D+1}\lambda _{\beta }^{\frac{D_1}{2}-1} F_{\beta
M}^{(1)N}(\lambda _{\beta },\lambda _{\beta })}{2^{D_1+1}\pi
^{\frac{D_1}{2}-1}z^{\frac{D_1}{2}}e^{2z\lambda _{\beta }}}
\frac{\bar{I}_{\nu }^{(a)}(z_a\lambda _{\beta })}{\bar{K}_{\nu
}^{(a)}(z_a\lambda _{\beta })} .
\end{equation}
Hence, the contribution of the nonzero KK modes exponentially
vanishes when the observation point tends to the AdS horizon, $z\to
\infty $. For the zero KK mode ($\lambda _{\beta }=0$) assuming
$z\gg z_a$, we see that the main contribution into the integral in
Eq. (\ref{EMT1bounda}) comes from $u$ for which $uz_{a}\ll 1$ and we
can replace the Bessel modified functions with this argument by
their asymptotic expressions for small values of the argument. The
remained integral is evaluated by the standard formula for the
integrals involving the square of the MacDonald function (see, for
instance, \cite{Prudnikov2}). For a homogeneous internal space to
the leading order this yields to the following formula
\begin{equation}\label{TMNsplsmallza2}
\langle T_{M}^{N}\rangle ^{(a)}_{\beta }\approx
-\frac{k_D^{D+1}z^{D_2}(D_1+2\nu )}{2^{D_1}\pi
^{\frac{D_1-1}{2}}c_{a}(\nu )} \frac{\Gamma \left(
\frac{D_1}{2}+2\nu \right) \Gamma \left( \frac{D_1}{2}+\nu \right)
}{\nu \Gamma ^2(\nu )\Gamma \left( \frac{D_1+1}{2}+\nu \right)}
\left( \frac{z_a}{2z}\right) ^{2\nu } F_{M}^{N},
\end{equation}
with the notations
\begin{subequations} \label{FMN}
\begin{eqnarray}
F_{\mu }^{\sigma }&=& \delta _{\mu }^{\sigma }\left[ \zeta
_{D_2}^{(-)} -\frac{D_1+4\nu }{4(D_1+2\nu +1)}\right] ,\\
F_{i}^{j}&=& \delta _{i}^{j}\zeta _{D_2}^{(-)} , \quad
F_{D}^{D}=\frac{D_1F_{0}^{0}+ F_{i}^{i}}{(D_1+2\nu )},
\end{eqnarray}
\end{subequations}
where
\begin{equation}\label{zetanpm}
\zeta _{n}^{(\pm )}=(n\pm 2\nu +1)\zeta -\frac{n\pm 2\nu }{4}.
\end{equation}
In particular, for a conformally coupled massless scalar one has
$F^{D}_{D}=0$. From (\ref{TMNsplsmallza2}) it follows that for the
zero mode the brane induced VEV near the AdS horizon behaves as
$z^{D_2-2\nu }$. In the purely AdS bulk ($D_2=0$) this VEV vanishes
on the horizon for $\nu
>0$. For an internal spaces with $D_2>2\nu $ the VEV diverges on the
horizon. Note that for a conformally coupled massless scalar and
$D_2=1$ the boundary induced VEV takes nonzero finite value on the
horizon. The VEV from the zero mode integrated over the internal
space (see Eq. (\ref{TMNintegrated})) vanishes on the AdS horizon
for all values $D_2$ due to the additional warp factor coming from
the volume element. In particular, the contribution of the brane
into the total vacuum energy per unit surface on the brane in the
region $[z,\infty ) $, $z>z_a$, is finite.

For large distances from the brane in the region $y<a$ one has $z\ll
z_a,1/\lambda _{\beta }$. The main contribution into the integral in
Eq. (\ref{Tik1plnewleft}) comes from the region $u\lesssim 1/z_a$ in
which $uz \ll 1$ and we can replace the function $I_{\nu }(uz)$ by
the corresponding expression for small values of the argument. As a
result to the leading order we obtain the formula
\begin{eqnarray}\label{TMNsplsmallz}
\langle T_{M}^{N}\rangle ^{(a)}_{\beta }&\approx &
\frac{2^{1-D_1-2\nu }k_D^{D+1}z^{D+2\nu }\zeta _{D}^{(+)}}{\pi
^{\frac{D_1}{2}}\Gamma \left( \frac{D_1}{2}\right) \Gamma ^2(\nu +1
)} \tilde F_{M}^{N} \nonumber
\\
&& \times \int_{\lambda _{\beta }}^{\infty }du\, u^{2\nu
+1}(u^2-\lambda _{\beta
}^2)^{\frac{D_1}{2}-1}\frac{\bar{K}_{\nu }^{(a)}(uz_{a})%
}{\bar{I}_{\nu }^{(a)}(uz_{a})} ,
\end{eqnarray}
with the coefficients defined by the relations
\begin{equation} \label{tildeFMN}
  \tilde F_{\mu }^{\sigma }=2\nu \delta _{\mu }^{\sigma },\;
  \tilde F_{i}^{k}=2\nu \delta _{i}^{k},\; \tilde F_{D}^{D}=-D,
  \; \tilde F_{D}^{i}=k_Dz^2 \eta _{\beta }^{i}(X).
\end{equation}
In this limit the brane induced vacuum stresses in the directions
parallel to the brane are isotropic. In the case of Dirichlet
boundary condition the corresponding energy density is negative for
both minimally and conformally coupled scalars. As we see for fixed
values $k_D$, $z_a$, $\lambda _{\beta }$ the brane induced VEV
vanishes as $z^{D+2\nu }$ for diagonal components and as $z^{D+2\nu
+2}$ for the ${}_D^i$--component when the observation point tends to
the AdS boundary. In particular, the contribution of the brane to
the total vacuum energy per unit surface on the brane in the region
$[0,z]$, $z<z_a$, is finite (near $z=0$ the integrand in the
corresponding integral over $z$ behaves as $z^{2\nu -1}$). The
integral in Eq. (\ref{TMNsplsmallz}) is simply estimated in two
subcases. For the case $z\ll \lambda _{\beta }^{-1}\ll z_a$ the main
contribution comes from the lower limit and to the leading order the
integral is proportional to $(\lambda _{\beta} /z_a)^{D_1/2}\lambda
_{\beta }^{2\nu }e^{-2\lambda _{\beta }z_a}$. As a result the brane
induced vacuum energy-momentum tensor is suppressed by the factor
$(\lambda _{\beta }z)^{D_1/2+2\nu }(z/z_a)^{D_1/2}e^{-2\lambda
_{\beta }z_a}$. In the opposite limit for $z_a$, $z\ll z_a\ll
\lambda _{\beta }^{-1}$, which corresponds to small KK masses,
$\lambda _{\beta }^{(a)}\ll k_D$, the lower limit in the integral
can be replaced by 0 and to the leading order the contribution of
the mode with a given $\beta $ does not depend on $\lambda _{\beta
}$ and behaves as $z^{D+2\nu }/z_a^{D_1+2\nu}$.

To see the convergence properties of the series over $\beta $ in Eq.
(\ref{TMNbetadef}), consider the contribution to the brane induced
VEV of the energy-momentum tensor from the modes with large KK
masses, $\lambda _{\beta }z,\lambda _{\beta }z_a \gg 1 $. Replacing
the Bessel modified functions by asymptotic expansions for large
values of the argument, to the leading order one finds
\begin{equation}\label{TMNspllargeKK}
\langle T_M^N\rangle ^{(a)}_{\beta }\approx
\frac{(k_Dz)^{D+1}}{(4\pi )^{\frac{D_1}{2}}\Gamma \left(
\frac{D_1}{2}\right) }\int _{\lambda _{\beta }}^{\infty } du\,
(u^2-\lambda _{\beta
}^2)^{\frac{D_1}{2}-1}\frac{e^{-2u|z-z_a|}}{c_a(uz_a)} F_{\beta
M}^{(1)N}(u,\lambda _{\beta }),
\end{equation}
with the notations defined by Eq. (\ref{FbetMNuv}),
\begin{equation}\label{cj1}
c_{j}(u)=\frac{ A_j-n^{(j)}B_j u}{A_j+n^{(j)}B_j u}, \quad j=a,b,
\end{equation}
and the definition for $n^{(j)}$ is given after formulae
(\ref{FbetMNuv}). Under the condition $\lambda _{\beta }z_a\gg
|A_a/B_a|$ or $B_a=0$, we have $c_a (uz_a)\approx \kappa (B_a)$ and
the integral can be expressed through the MacDonald function by
using the formula
\begin{equation}\label{intu}
\int _{\lambda _{\beta }}^{\infty } du\, u^n (u^2-\lambda _{\beta
}^2)^{\frac{D_1}{2}-1}e^{-2u\eta }=\frac{\lambda _{\beta
}^n}{\sqrt{\pi }}\Gamma \left( \frac{D_1}{2}\right)\left(
\frac{\lambda _{\beta }}{\eta }\right)
^{\frac{D_1-1}{2}}K_{\frac{D_1-1}{2}+n}(2\lambda _{\beta }\eta ) ,
\end{equation}
where $n=0,1$. Additionally assuming $\lambda _{\beta }|z-z_a|\gg
1$, from (\ref{TMNspllargeKK}) we find that the contribution of KK
modes with large masses to the leading order is given by
\begin{equation}\label{TMNspllargeKK1}
\langle T_M^N\rangle ^{(a)}_{\beta }\approx
\frac{(k_Dz)^{D+1}\lambda _{\beta }^{\frac{D_1}{2}-1}e^{-2\lambda
_{\beta }|z-z_a|}}{2^{D_1+1}\pi ^{\frac{D_1}{2}}c_{a}(\lambda
_{\beta }z_a)|z-z_a|^{\frac{D_1}{2}}}F_{\beta M}^{(1)N}(\lambda
_{\beta },\lambda _{\beta }) ,
\end{equation}
and is exponentially suppressed. In particular, the corresponding
conditions are satisfied for all nonzero KK modes if the length
scale of the internal space $\Sigma $ is sufficiently small. In this
case the main contribution into the vacuum energy-momentum tensor
comes from the zero mode. In the opposite limit, when the length
scale of the internal space is large, to the leading order the
vacuum energy-momentum tensor reduces to the corresponding result
for a brane in the bulk $AdS_{D+1}$ given in Ref. \cite{Saha04a}.

In the limit of strong gravitational fields, corresponding to large
values of the AdS energy scale $k_D$ when the values of the other
parameters are fixed, for nonzero KK modes along $\Sigma $ one has
$\lambda _{\beta }z_a, \lambda _{\beta }z\gg 1$. Hence, the behavior
of the brane induced VEV of the energy-momentum tensor in this case
can be estimated by formula (\ref{TMNspllargeKK}). If in addition
the condition $\lambda _{\beta }|z-z_a|\gg 1$ is satisfied, we have
formula (\ref{TMNspllargeKK1}) with the exponential suppression of
the brane induced part. For the zero KK mode the corresponding
behavior of the brane induced vacuum energy-momentum tensor is
estimated by using the asymptotic formulae for the Bessel modified
functions for small and large values of the argument in the regions
$y>a$ and $y<a$ respectively. To the leading order the components of
this tensor behave like $k_{D}^{D_1+1}e^{D_2k_Dy}\exp [(D_1+2\nu
)k_D(y-a)]$ in the region $y<a$ and like
$k_{D}^{D_1+1}e^{D_2k_Dy}\exp [2\nu k_D(a-y)]$ in the region $y>a$.
The corresponding quantities integrated over the internal space
contain additional factor $e^{-D_2k_Dy}$ coming from the volume
element and are exponentially small in both regions.

Now we turn to the consideration of the limiting cases for the
parameter $z_a$ determining the position of the brane. For small
values of this parameter, $z_a\ll z,1/\lambda _{\beta }$, for the
dominant contribution into the $u$-integral in Eq.
(\ref{EMT1bounda}) one has $u z_a\ll 1$ and, using the asymptotic
formulae for the Bessel modified functions for small values of the
argument, one finds
\begin{equation}\label{TMNsplsmallza}
\langle T_{M}^{N}\rangle ^{(a)}_{\beta }\approx -\frac{2^{2-D_1-2\nu
}k_D^{D+1}z^Dz_a^{2\nu }}{\pi ^{\frac{D_1}{2}}\Gamma \left(
\frac{D_1}{2}\right) \nu \Gamma ^2(\nu )c_a(\nu )} \int_{\lambda
_{\beta }}^{\infty }du\, u^{2\nu +1}(u^2-\lambda _{\beta
}^2)^{\frac{D_1}{2}-1}F_{\beta M}^{(+)N}[K_{\nu }(uz)].
\end{equation}
Note that in this limit the distance of the observation point from
the brane is large compared with the AdS curvature radius,
$k_D(y-a)\gg 1$, and the KK masses $\lambda _{\beta }^{(a)}$ are
small with respect to the AdS energy scale, $\lambda _{\beta
}^{(a)}\ll k_D$. In particular, from Eq. (\ref{TMNsplsmallza}) it
follows that for fixed values $k_D$, $y$, $\lambda _{\beta }$ the
brane induced VEV in the region $z>z_a$ vanishes as $z_a^{2 \nu }$
when the brane position tends to the AdS boundary, $z_a\to 0$.
Formula (\ref{TMNsplsmallza}) is further simplified for two
subcases. In the case $z_a\ll \lambda _{\beta }^{-1}\ll z$, or
equivalently $k_De^{-k_D(y-a)}\ll \lambda _{\beta }^{(a)}\ll k_D$,
the main contribution into the integral comes from the lower limit
ant to the leading order we find
\begin{equation}\label{phi2splsmallza1}
\langle T_{M}^{N}\rangle ^{(a)}_{\beta }\approx
\frac{k_D^{D+1}z^{D_2}(z_a\lambda _{\beta })^{2\nu }(z\lambda
_{\beta })^{\frac{D_1}{2}+1}}{2^{D_1+2\nu }\pi ^{\frac{D_1}{2}-1}\nu
\Gamma ^2(\nu )c_{a}(\nu )\lambda _{\beta }^2} e^{-2z\lambda _{\beta
}}F_{\beta M}^{(1)N}(\lambda _{\beta },\lambda _{\beta }).
\end{equation}
In the limit $z_a\ll z\ll \lambda _{\beta }^{-1}$ (small KK masses,
$\lambda _{\beta }^{(a)}\ll k_De^{-k_D(y-a)}$) to the leading order
we can put 0 in the lower limit of the integral and we obtain
formula (\ref{TMNsplsmallza2}).

When the brane position tends to the AdS horizon, $z_a\to \infty $,
for massive KK modes along $\Sigma $ the main contribution into the
VEV of the energy-momentum tensor in the region $z<z_a$ comes from
the lower limit of the $u$-integral. To the leading order we find
\begin{equation}\label{TMNsplzanearhor}
\langle T_{M}^{N}\rangle ^{(a)}_{\beta }\approx - \frac{k_D^{D+1}z^D
\lambda _{\beta }^{\frac{D_1}{2}}e^{-2\lambda _{\beta
}z_a}}{2^{D_1}\pi ^{\frac{D_1}{2}-1}z_a^{D_1/2}} \frac{F_{\beta
M}^{(+)N}[I_{\nu }(\lambda _{\beta } z)]}{c_a(\lambda _{\beta }z_a)}
,
\end{equation}
and the VEV is exponentially small. For the zero mode in the same
limit, introducing a new integration variable $v=uz_a$, we expand
the functions $F_{\beta M}^{(+)N}[I_{\nu }(vz/z_a)]$ for small
values of the argument. To the leading order this yields
\begin{equation}\label{TMNspllargeza0mode}
\langle T_{M}^{N}\rangle ^{(a)}_{\beta }\approx \frac{k_{D}^{D+1}
z^{D_2}\left[ \zeta _{D}^{(+)}\tilde F_{ M}^{N}+z^2F_{\beta
M}^{(1)N}(0,0)\right] }{2^{D_1+2\nu }\pi ^{\frac{D_1}{2}}\Gamma
\left( \frac{D_1}{2}\right) \Gamma ^{2}(\nu +1)} \left(
\frac{z}{z_a}\right) ^{D_1+2\nu }\int_{0}^{\infty }du\, u^{D_1+2\nu
-1} \frac{\bar{K}_{\nu }^{(a)}(u)}{\bar{I}_{\nu }^{(a)}(u)} \, ,
\end{equation}
and the suppression is power-law with respect to $z_a$.

\section{Energy-momentum tensor in the region between two branes }

\label{sec:EMT2pl}

In the geometry of two branes we have three distinct regions of the
radial coordinate: $y<a$, $a<y<b$, and $y>b$. The VEVs in the first
and last regions are the same as those for a single brane. In this
section we will discuss the vacuum energy-momentum tensor in the
region between two branes. Note that in the orbifolded version of
the model this region is employed only. By using formula
(\ref{vevEMT1pl}), the VEV of the energy-momentum tensor is
expressed through the corresponding Wightman function and the VEV of
the field square investigated in Ref. \cite{Saha05I}. The Wightman
function is presented in the form
\begin{eqnarray}
\langle 0|\varphi (x)\varphi (x^{\prime })|0\rangle &=&\langle
\varphi (x)\varphi (x^{\prime })\rangle ^{(0)} +\langle \varphi
(x)\varphi (x^{\prime
})\rangle ^{(j)}-\frac{k_{D}^{D-1}(zz^{\prime })^{\frac{D}{2}}}{2^{D_{1}-1}\pi ^{D_{1}}}%
\sum_{\beta }\psi _{\beta }(X)\psi _{\beta }^{\ast }(X^{\prime })  \nonumber \\
&\times & \int d{\bf k%
}\,e^{i{\bf k}({\bf x}-{\bf x}^{\prime })} \int_{k_{\beta }}^{\infty
}du u G_{\nu
}^{(j)}(uz_{j},uz)G_{\nu }^{(j)}(uz_{j},uz^{\prime }) \nonumber \\
&\times & \frac{\Omega _{j\nu }(uz_{a},uz_{b})}{\sqrt{u^{2}-k
_{\beta }^{2}}} \cosh \left[ \sqrt{u^{2}-k_{\beta }^{2}}%
(t-t^{\prime })\right] ,  \label{W15}
\end{eqnarray}
where $k_{\beta }$ is defined after formula (\ref{WF1bran}) and
$j=a,b$ provide two equivalent representations. Here and in the
formulae below we use the notations
\begin{subequations} \label{Omn}
\begin{eqnarray}
\Omega _{a\nu }(u,v) &=&\frac{\bar{K}_{\nu }^{(b)}(v)/\bar{K}_{\nu }^{(a)}(u)%
}{\bar{K}_{\nu }^{(a)}(u)\bar{I}_{\nu }^{(b)}(v)-\bar{K}_{\nu }^{(b)}(v)\bar{%
I}_{\nu }^{(a)}(u)},  \label{Omnu} \\
\Omega _{b\nu }(u,v)&=&\frac{\bar{I}_{\nu }^{(a)}(u)/\bar{I}_{\nu }^{(b)}(v)}{%
\bar{K}_{\nu }^{(a)}(u)\bar{I}_{\nu }^{(b)}(v)-\bar{K}_{\nu }^{(b)}(v)\bar{I}%
_{\nu }^{(a)}(u)},  \label{Omnub}
\end{eqnarray}
\end{subequations}
and
\begin{equation}
G_{\nu }^{(j)}(u,v) =I_{\nu }(v)\bar{K}_{\nu }^{(j)}(u)-\bar{I}_{\nu
}^{(j)}(u)K_{\nu }(v).  \label{Geab}
\end{equation}%
Substituting the Wightman function into formula (\ref{vevEMT1pl})
and using relation (\ref{relintk}), for the components of the vacuum
energy-momentum tensor in the region between the branes we obtain
the formula
\begin{eqnarray}
\langle 0|T_{M}^{N}|0\rangle &=&\langle T_{M}^{N}\rangle
^{(0)}+\langle T_{M}^{N}\rangle ^{(j)}- \frac{2k_D^{D+1}z^{D}}{(4\pi
) ^{\frac{D_1}{2}} \Gamma \left(
\frac{D_1}{2}\right) }\sum_{\beta }|\psi _{\beta }(X)|^2\nonumber \\
&& \times \int _{\lambda _{\beta }}^{\infty } d u\, u
 (u^2-\lambda _{\beta }^2)^{\frac{D_1}{2}-1} \Omega _{j\nu }(u z_{a},u z_{b})
 F_{\beta M}^{(+)N}[G^{(j)}_{\nu }(uz_j,uz)],
\label{Tikjint}
\end{eqnarray}
with the functions $F_{\beta M}^{(+)N}[g(v)]$ defined by relations
(\ref{Fmu})--(\ref{Fi}), where $g(v)=G_{\nu }^{(j)}(uz_j,v)$, and
with $j=a,b$. The last term on the right of this formula is finite
on the brane at $z=z_j$ and diverges for the points on the brane
$z=z_{j'}$, $j'=a,b$, $j'\neq j$. These divergences are the same as
those for a single brane at $z=z_{j'}$. As a result, if we write the
VEV of the energy-momentum tensor in the form
\begin{equation}\label{TMNtwopl1}
    \langle 0|T_{M}^{N}|0\rangle = \langle T_{M}^{N}\rangle ^{(0)}+
    \sum_{j=a,b}\langle T_{M}^{N}\rangle
^{(j)}+\langle T_{M}^{N}\rangle ^{(ab)},
\end{equation}
then the interference part $\langle T_{M}^{N}\rangle ^{(ab)}$ is
finite on both branes. In the case of one parameter manifold $\Sigma
$ with the size $L$ and for a given $k_D$, the VEV in Eq.
(\ref{TMNtwopl1}) is a function on $z_b/z_a$, $L/z_a$, and $z/z_a$.
By using formula (\ref{Tikjint}) and the expression for a single
brane induced VEV of the energy-momentum tensor, we can present the
contribution of the given KK mode along $\Sigma $ into the
interference part of Eq. (\ref{TMNtwopl1}) in the form
\begin{eqnarray} \label{intpart}
\langle T_{M}^{N}\rangle ^{(ab)}_{\beta } &=&
\frac{2k_D^{D+1}z^{D}}{(4\pi )^{\frac{D_1}{2}} \Gamma \left(
\frac{D_1}{2}\right) } \int _{\lambda _{\beta }}^{\infty } d u
\frac{u
 (u^2-\lambda _{\beta }^2)^{\frac{D_1}{2}-1}}{\frac{\bar{K}_{\nu }^{(a)}(uz_a)
 \bar{I}_{\nu }^{(b)}(u z_b)}{\bar{I}_{\nu }^{(a)}(u z_a)\bar{K}_{\nu }^{(b)}(u z_b)}
 -1} \bigg\{ 2F_{\beta M}^{(+)N}[I_{\nu }(u z),K_{\nu }(u z)] \nonumber \\
  && - \frac{\bar{I}_{\nu }^{(a)}(u z_a)}{\bar{K}_{\nu }^{(a)}(u z_a)}
  F_{\beta M}^{(+)N}[K_{\nu }(u z)]- \frac{\bar{K}_{\nu }^{(b)}(u z_b)}{\bar{I}_{\nu }^{(b)}
  (u z_b)} F_{\beta M}^{(+)N}[I_{\nu }(u z)] \bigg\} ,
\end{eqnarray}
where $\langle T_{M}^{N}\rangle ^{(ab)}_{\beta }$ is defined by the
relation similar to Eq. (\ref{TMNbetadef}), and the functions
$F_{\beta M}^{(+)N}[g_1,g_{2}]$ are defined by formulae
(\ref{Fmu})--(\ref{Fi}) with the replacements $g^{\prime 2}\to
g^{\prime }_1 g^{\prime }_2$, $gg^{\prime }\to (g_1g_2)^{\prime
}/2$, and $g^2\to g_1g_2$. For large values $u$ the integrand in
formula (\ref{intpart}) behaves as $e^{-2u(z_b-z_a)}$ and the
integral is convergent for all values $z$ including the points on
the branes. For interbrane distances much smaller than the AdS
curvature radius, $(b-a)\ll 1/k_D$, the main contribution into the
integral in Eq. (\ref{intpart}) comes from large values $u$ and we
can replace the Bessel modified functions by asymptotic expressions
for large values of the argument. Additionally assuming that the
distance is small compared with the KK length scale for an observer
living on the brane $y=a$, $(b-a)\ll 1/\lambda _{\beta }^{(a)}$, it
can be seen that the diagonal components of the interference part
with $M\neq D$ behave as $(b-a)^{-D_1-1}$, whereas ${}^{D}_{D}$--
and ${}^{i}_{D}$--components behave like $(b-a)^{-D_1}$.

Now let us consider various limiting cases when the general formula
for the interference part is simplified. First of all, in the limit
$k_D\to 0$ the order of Bessel modified functions is large and we
use the uniform asymptotic expansions for these functions. To the
leading order the following result is obtained
\begin{eqnarray}\label{TMNtwoplMink}
 \langle T_{M}^{N}\rangle ^{(ab)}& \approx & \frac{(4\pi )^{-
    \frac{D_1}{2}}}{\Gamma \left( \frac{D_1}{2}\right)}\sum_{\beta
}\left| \psi _{\beta }(X)\right| ^2 \int_{v_{\beta }}^{\infty } du\,
\frac{(u^2-v_{\beta }^{2})^{\frac{D_1}{2}-1}}{\tilde
c_{a}(u a)\tilde c_{b}(u b)e^{2u(b-a)}-1} \nonumber \\
&& \times \bigg[ \sum_{j=a,b}\frac{e^{-2u|y-j|}}{\tilde
c_{j}(u)}F_{j\beta M}^{(1)N}(u,v_{\beta })-2F_{\beta
M}^{(0)N}(u,v_{\beta })\bigg] _{k_D=0} ,
\end{eqnarray}
with notations defined by Eq. (\ref{FbetMNuv}) and $v_{\beta
}=\sqrt{m^2+ \lambda _{\beta }^{2}}$. In the expression on the right
the functions $F_{j\beta M}^{(1)N}(u,v)$ are defined by the same
relations as the functions $F_{\beta M}^{(1)N}(u,v)$ from
(\ref{FbetMNuv}) with the replacement $n^{(a)}\to n^{(j)}$ in
formula (\ref{FbetMNuv3}), and with $n^{(j)}$ defined in the
paragraph after formula (\ref{FbetMNuv}). This limit corresponds to
the geometry of two parallel Robin plates in the bulk
$R^{(D_1,1)}\times \Sigma $ and the expression on the right of
formula (\ref{TMNtwoplMink}) coincides with the interference part
$\langle T_{M}^{N}\rangle ^{(ab)}_{R^{(D_1,1)}\times \Sigma }$ in
the VEV of the energy-momentum tensor in the region between two
plates. This expression generalizes the corresponding result from
Ref.~ \cite{Rome02} for Robin plates in the Minkowski background.

For large KK masses, $z_{a}\lambda _{\beta }\gg 1$, the arguments of
the Bessel modified functions in Eq. (\ref{intpart}) are large. By
using the corresponding asymptotic expansions, for the contribution
of the mode with a given $\beta $ one finds
\begin{eqnarray}\label{TMN2pllargelamb}
\langle T_{M}^{N}\rangle ^{(ab)}_{\beta }& \approx &
\frac{(k_Dz)^{D+1}}{(4\pi )^{\frac{D_1}{2}}\Gamma \left(
\frac{D_1}{2}\right)} \int_{\lambda _{\beta }}^{\infty } du\,
\frac{(u^2-\lambda _{\beta }^{2})^{\frac{D_1}{2}-1}}{c_{a}(u
z_a) c_{b}(u z_b)e^{2u(z_b-z_a)}-1} \nonumber \\
&& \times \bigg[ \sum_{j=a,b}\frac{e^{-2u|z-z_j|}}{c_{j}(u
z_j)}F_{j\beta M}^{(1)N}(u,\lambda _\beta )-2F_{\beta M}^{(0)
N}(u,\lambda _\beta )\bigg] .
\end{eqnarray}
In particular, for a one parameter internal manifold with the length
scale $L\ll z_a$, this formula is valid for all nonzero KK modes
along $\Sigma $. Under the additional assumption $\lambda _{\beta
}|z-z_j|\gg 1$, the main contribution into the integral in this
formula comes from the lower limit and we obtain the formula
\begin{equation}\label{TMN2pllargelamb1}
\langle T_{M}^{N}\rangle ^{(ab)}_{\beta }\approx
-\frac{(k_Dz)^{D+1}}{(4\pi )^{\frac{D_1}{2}}} \frac{\lambda _{\beta
}^{\frac{D_1}{2}-1}e^{-2\lambda _{\beta }(z_b-z_a)}F_{\beta
M}^{(0)N}(\lambda _\beta ,\lambda _\beta )}{c_{a}(\lambda _{\beta
}z_a)c_{b}(\lambda _{\beta }z_b)(z_b-z_a)^{\frac{D_1}{2}}} .
\end{equation}
In particular, it follows from here that in the expressions for the
brane induced VEVs the series over KK modes along $\Sigma $ are
exponentially convergent. Comparing with (\ref{TMNspllargeKK1}) we
see that for large KK masses the interference part is exponentially
suppressed with respect to the single brane parts.

Now let us assume the conditions $\lambda _{\beta }z_b\gg 1$ and
$z_a \lambda _{\beta }\lesssim 1$. This limit corresponds to large
interbrane distances compared with the AdS curvature radius
$k_D^{-1}$ and is realized in braneworld scenarios for the solution
of the hierarchy problem. In this limit the main contribution into
the $u$-integral comes from the region near the lower limit and for
the interference part to the leading order we have
\begin{eqnarray}
 \langle T_{M}^{N}\rangle ^{(ab)}_{\beta }&\approx & \frac{k_D^{D+1}
 z^{D}\lambda _{\beta }^{\frac{D_1}{2}}e^{-2\lambda _{\beta }z_b}}{2^{D_1}
 \pi ^{\frac{D_1}{2}-1}z_b^{\frac{D_1}{2}}c_{b}(\lambda _{\beta }z_b)}
 \frac{\bar{I}_{\nu }^{(a)}(z_a\lambda _{\beta })
}{\bar{K}_{\nu }^{(a)}(z_a\lambda _{\beta })} \nonumber \\
   & & \times \bigg\{
2F_{\beta M}^{(+)N}[I_{\nu }(\lambda _{\beta }z),K_{\nu }(\lambda
_{\beta }z)] -\frac{\bar{I}_{\nu }^{(a)}(z_a\lambda _{\beta
})}{\bar{K}_{\nu }^{(a)}(z_a\lambda _{\beta })}F_{\beta
M}^{(+)N}[K_{\nu }(\lambda _{\beta }z)]\bigg\} ,
\label{TMN2pllargezbn}
\end{eqnarray}
for the nonzero KK modes along $\Sigma $. In the limit $z\ll \lambda
_{\beta }^{-1}\ll z_b$ after the replacement of the Bessel modified
functions by their asymptotic expressions for small values of the
argument, the corresponding formula takes the form
\begin{eqnarray}\label{TMN2plas1a}
\langle T_{M}^{N}\rangle ^{(ab)}_{\beta }& \approx &
\frac{k_D^{D+1}z^{D_2}(z_a\lambda _{\beta })^{2\nu }(z\lambda
_{\beta })^{\frac{D_1}{2}}e^{-2\lambda _{\beta }z_b}(z/z_{b})
^{\frac{D_1}{2}}}{2^{D_1+2\nu -1}\pi ^{\frac{D_1}{2}-1}c_{a}(\nu
)c_{b}(z_b\lambda _{\beta })\Gamma ^2(\nu +1)} \nonumber
\\
&& \times \bigg[ \frac{m^2}{k_D^2}\delta _{M}^{D}\delta
_{D}^{N}+\frac{\zeta _{D}^{(+)}\tilde F_{M}^{N}}{2 c_a(\nu )}\left(
\frac{z_a}{z}\right) ^{2\nu }\bigg] ,
\end{eqnarray}
where $\tilde F_{M}^{N} $ is defined by relations (\ref{tildeFMN}).
For $\lambda _{\beta }z\gg 1$ and $\lambda _{\beta }z_a \lesssim 1$,
using the asymptotic formulae for the Bessel modified functions for
large values of the argument, we obtain the following formula
\begin{eqnarray} \label{TMN2pllargez1}
\langle T_{M}^{N}\rangle ^{(ab)}_{\beta }& \approx &
-\frac{(k_Dz)^{D+1} \lambda _{\beta }^{\frac{D_1}{2}-1}e^{-2\lambda
_{\beta }z_b}}{2^{D_1}\pi ^{\frac{D_1}{2}-1}c_{b}(\lambda _{\beta
}z_b)z_b^{D_1/2}} \frac{\bar{I}_{\nu }^{(a)}(z_a\lambda
_{\beta })}{\bar{K}_{\nu }^{(a)}(z_a\lambda _{\beta })} \nonumber \\
 && \times \bigg[ F_{\beta M}^{(0)N}(\lambda _{\beta }, \lambda _{\beta })-
 \frac{e^{-2\lambda _{\beta }(z_b-z)}}{2c_{b}(\lambda _{\beta }z_b)}
 F_{b\beta M}^{(1)N}(\lambda _{\beta }, \lambda _{\beta })\bigg] .
\end{eqnarray}
In the limit $z_a \ll z_b\ll \lambda _{\beta }^{-1}$, to the leading
order we can put 0 instead of $\lambda _{\beta }$ in the lower limit
of the integral over $u$, and by using the asymptotic formulae for
the Bessel modified functions for small values of the argument, it
can be seen that $\langle T_M^N\rangle ^{(ab)}_{\beta }\sim
(z_a/z_b)^{2\nu } g(z/z_b)$. In particular, it follows from here
that the interference part in the vacuum energy-momentum tensor
vanishes as $z_a^{2\nu }$ when the left brane tends to the AdS
boundary, $z_a\to 0$. Under the condition $z\ll z_b$ an additional
suppression factor appears in the form $(z/z_b)^{D_1}$ for
${}^{D}_{D}$--component and in the form $(z/z_b)^{D_1+2\alpha _1}$
for the other components, where $\alpha _{1}=\min (1,\nu )$. As we
see for large interbrane distances the interference part of the
brane induced VEV of the energy-momentum tensor is mainly located
near the brane $z=z_b$.

In the higher dimensional generalization of the Randall-Sundrum
braneworld based on the bulk $AdS_{D_1+1}\times \Sigma $, $y$
coordinate is compactified on an orbifold $S^{1}/Z_2$ and the
orbifold fixed points are the locations of two $D$-dimensional
branes. The corresponding VEVs of the energy-momentum tensor are
determined by the formulae given in this section with an additional
factor 1/2 and with Robin coefficients (see also \cite{Saha05I})
\begin{equation}\label{AtildeRS}
\frac{\tilde A_a}{\tilde B_a} =  -\frac{1}{2}(c_1+4D\zeta k_D),\quad
\frac{\tilde A_b}{\tilde B_b} = -\frac{1}{2}(-c_2+4D\zeta k_D),
\end{equation}
for untwisted scalar field. Here $c_1$ and $c_2$ are surface mass
parameters on the branes $y=a$ and $y=b$, respectively. For twisted
scalar Dirichlet boundary conditions are obtained. The one-loop
effective potential and the problem of moduli stabilization in this
model with zero mass parameters $c_j$ are discussed in Ref.
\cite{Flac03b}. A scenario is proposed where supersymmetry is broken
near the fundamental Planck scale, and the hierarchy between the
electroweak and effective Planck scales is generated by a
combination of redshift and large volume effects. The corresponding
parameters satisfy the relations  $L\lesssim 1/k_D$, $z_a\sim L$,
and $z_b/z_a \gg 1$. In this case the contribution into the
interference part of the energy-momentum tensor from the nonzero KK
modes is estimated by formula (\ref{TMN2pllargezbn}). For the zero
mode the interference part behaves like $e^{2\nu k_D(a-b)}$ and is
exponentially small.

\section{Interaction forces}

\label{sec:Intforce}

Now we turn to the vacuum forces acting on the branes. The
corresponding effective pressure $p^{(j)}$ acting on the brane at
$z=z_j$ is determined by ${}^D_D$--component of the vacuum
energy-momentum tensor evaluated at the point of the brane location:
$p^{(j)}=-\langle T^{D}_{D}\rangle _{z=z_j}$. For the region between
two branes it can be presented as a sum of two terms:
\begin{equation}
p^{(j)}=p^{(j)}_1+p^{(j)}_{{\mathrm{(int)}}},\quad j=a,b.
\label{pintdef}
\end{equation}
The first term on the right is the pressure for a single brane at
$z=z_j$ when the second brane is absent. This term is divergent due
to the surface divergences in the VEVs and needs additional
renormalization. This can be done, for example, by applying the
generalized zeta function technique to the corresponding mode sum.
This procedure is similar to that used in Ref. \cite{Flac03b} for
the evaluation of the effective potential. The corresponding
calculation lies on the same line with the evaluation of the surface
Casimir densities and will be presented in the forthcoming paper.
Here we note that the single brane term in the vacuum effective
pressure will contain a part which depends on the renormalization
scale. This part will change under the change of the renormalization
scale and can be fixed by imposing suitable renormalization
conditions which relates it to observables. Below we will be
concentrated on the second term in the right of Eq. (\ref{pintdef}).
This term is the additional vacuum pressure induced by the presence
of the second brane, and can be termed as an interaction force. It
is determined by the last term on the right of formulae
(\ref{Tikjint}) evaluated at the brane location $z=z_j$. It is
finite for all nonzero interbrane distances and is not changed by
the regularization and renormalization procedure. In particular, it
does not depend on the type of the regularization procedure used.
All regularization umbiguities are involved in $p_{1}^{(j)}$.
Substituting $z=z_j$ into the second term on the right of formula
(\ref{Tikjint}) and using the relations
\begin{equation}
G_{\nu }^{(j)}(u,u)=-B_j,\quad \frac{dG_{\nu }^{(j) }(u,v)}{dv}\Big|
_{v=u}=\frac{A_j}{u}, \label{Guu}
\end{equation}
for the interaction part of the vacuum effective pressure one has
\begin{equation}
p^{(j)}_{{\mathrm{(int)}}}= \frac{k_D^{D+1}z^{D}_{j}}{(4\pi )
^{\frac{D_1}{2}} \Gamma \left( \frac{D_1}{2}\right) }\sum_{\beta
}|\psi _{\beta }(X)|^2\int _{\lambda _{\beta }}^{\infty } d u u
 (u^2-\lambda _{\beta }^2)^{\frac{D_1}{2}-1} \Omega _{j\nu }(u z_{a},u z_{b})
 F_{\beta }^{(j)}(u z_j), \label{pintj}
\end{equation}
where we have introduced the notation
\begin{equation}\label{Fbetj}
 F_{\beta }^{(j)}(u)=\left( u^2
 -\nu ^2 +2\frac{m^2}{k_D^2}\right) B_j^2 -
D(4\zeta -1) A_jB_j-A_j^2 -2\left( \zeta -\frac{1}{4}
  \right) z_j^2B_j^2\eta _{\beta }(X) .
\end{equation}
Below we will show that for small interbrane distances the
interaction part dominates the single brane parts in Eq.
(\ref{pintdef}). For a Dirichlet scalar $B_j=0$ and one has
$F_{\beta }^{(j)}(u)=-A_j^2$. By using the properties of the
modified Bessel function it can be seen that in this case $\Omega
_{j\nu }(u z_{a},u z_{b})>0$ and, hence, the vacuum interaction
forces are attractive. For a given value of the AdS energy scale
$k_D$ and one parameter manifold $\Sigma $ with size $L$, the vacuum
interaction forces (\ref{pintj}) are functions on the ratios
$z_b/z_a$ and $L/z_a$. The first ratio is related to the proper
distance between the branes and the second one is the ratio of the
size of the internal space measured by an observer residing on the
brane at $y=a$ to the AdS curvature radius $k_D^{-1}$. Note that the
term 'interaction' in the discussion here and below should be
understood conditionally. The quantity $p_{\mathrm{(int)}}^{(j)}$
determines the force by which the scalar vacuum acts on the brane
due to the modification of the spectrum for the zero-point
fluctuations by the presence of the second brane. As the vacuum
properties depend on the coordinate $y$, there is no a priori reason
for the interaction terms (and also for the total pressures
$p^{(j)}$) to be equal for the branes $j=a$ and $j=b$, and the
corresponding forces in general are different even in the case of
the same Robin coefficients in the boundary conditions.

Now we turn to the limiting cases when the expressions for the
interaction forces between the branes are simplified. First of all
we consider the limit $k_D\to 0$. By the way similar to that used
before for the vacuum energy-momentum tensor, to the leading order
we find
\begin{eqnarray}
 p^{(j)}_{{\mathrm{(int)}}}&\approx & -2\frac{(4\pi )^{-\frac{D_1}{2}}}{
 \Gamma \left( \frac{D_1}{2}\right)} \sum_{\beta
}|\psi _{\beta }(X)|^2\int _{v_{\beta }}^{\infty } du\,
\frac{u^2(u^2-v_{\beta }^2)^{\frac{D_1}{2}-1}}{\tilde c_{a}(u)
\tilde c_{b}(u)e^{2u(b-a)}-1} \nonumber \\
  && \times \left[ 1+\frac{(2\zeta -1/2)\tilde B_j^2}{\tilde A_j^2-u^2\tilde B_j^2}
  \eta _{\beta }(X) \right] , \label{pintkD0}
\end{eqnarray}
where $v_{\beta }$ is defined after formula (\ref{TMNsplMink}). The
expression on the right of this formula presents the corresponding
force acting per unit surface on the brane in the bulk geometry
$R^{(D_1-1,1)}\times \Sigma $. Note that in this case the vacuum
effective pressures are the same for both branes if the coefficients
in the boundary conditions are the same. Moreover, for a homogeneous
internal space, the contribution of the second term in the square
brackets on the right of Eq. (\ref{pintkD0}) vanishes, and the
interaction forces are the same even in the case of different Robin
coefficients for separate branes.

For large values of KK masses along $\Sigma $, $\lambda _{\beta
}z_j\gg 1$, we can replace Bessel modified functions by their
asymptotic expansions for large values of the argument. For the
contribution of a given KK mode to the leading order this gives
\begin{eqnarray}
 p^{(j)}_{{\mathrm{(int)}}\beta }&\approx & -\frac{2(k_Dz_j)^{D+1}}{(4\pi )^{\frac{D_1}{2}}
 \Gamma \left( \frac{D_1}{2}\right)} \int _{\lambda _{\beta }}^{\infty } du\,
\frac{u^2(u^2-\lambda _{\beta }^2)^{\frac{D_1}{2}-1}}{c_{a}(uz_a)
c_{b}(uz_b)e^{2u(z_b-z_a)}-1} \nonumber \\
  && \times \left[ 1+\frac{(2\zeta -1/2)z_j^2 B_j^2}{A_j^2-(uz_j B_j)^2}
  \eta _{\beta }(X)
  \right] , \label{pintlargeKK}
\end{eqnarray}
where $p^{(j)}_{{\mathrm{(int)}}\beta }$ is determined by the
relation similar to Eq. (\ref{TMNbetadef}). If in addition one has
the condition $\lambda _{\beta }(z_b-z_a)\gg 1$, the main
contribution into the $u$-integral comes from the lower limit and we
have the formula
\begin{eqnarray}
 p^{(j)}_{{\mathrm{(int)}}\beta }&\approx & -\frac{(k_Dz_j)^{D+1}}{(4\pi )^{\frac{D_1}{2}}
 }\frac{\lambda _{\beta }^{D_1/2+1}e^{-2\lambda _{\beta }(z_b-
 z_a)}}{c_{a}(\lambda _{\beta }z_a) c_{b}(\lambda _{\beta }z_b)(z_b-z_a)^{\frac{D_1}{2}}} \nonumber \\
  && \times \left[ 1+\frac{(2\zeta -1/2)z_j^2 B_j^2}{A_j^2-(\lambda _{\beta }z_j B_j)^2}
  \eta _{\beta }(X)\right] . \label{pintlargeKK1}
\end{eqnarray}
In particular, for sufficiently small length scale of the internal
space this formula is valid for all nonzero KK masses and the main
contribution to the interaction forces comes from the zero KK mode.
In the opposite limit of large internal space, to the leading order
we obtain the corresponding result for parallel branes in
$AdS_{D+1}$ bulk \cite{Saha04a}.

For small interbrane distances, $k_D(b-a)\ll 1$, which is equivalent
to $z_b/z_a-1\ll 1$, the main contribution into the integral in Eq.
(\ref{pintj}) comes from large values $u$ and to the leading order
we obtain formula (\ref{pintlargeKK}). If in addition one has
$\lambda _{\beta }(z_b-z_a)\ll 1$ or equivalently $\lambda _{\beta
}^{(a)}(b-a)\ll 1$, and assuming $(b-a)\ll |\tilde B_j/A_j|$ or
$\tilde B_j=0$, we can put in this formula $\lambda _{\beta }=0$,
and to the leading order one finds
\begin{equation}\label{pintsmalldist}
p^{(j)}_{{\mathrm{(int)}}\beta }\approx a_{D_1}\frac{D_1\Gamma
\left( \frac{D_1+1}{2}\right) \zeta _{R}(D_1+1)}{(4\pi )
^{(D_1+1)/2}(b-a)^{D_1+1}}e^{D_2k_Dj},
\end{equation}
where $a_{D_1}=1-2^{-D_1}$ for $\kappa (B_a) \kappa (B_b)=-1$ and
$a_{D_1}=-1$ for $\kappa (B_a) \kappa (B_b)=1$. It follows from here
that for small interbrane distances the interaction forces are
repulsive for Dirichlet boundary condition on one brane and
non-Dirichlet boundary condition on the another and are attractive
for other cases. As in the limit $a\to b$ the renormalized values of
the single brane parts $p^{(j)}_1$, $j=a,b$, are finite, for small
interbrane distances the main contribution into the vacuum effective
pressure $p^{(j)}$ comes from the interaction part.

Now we consider the limit $\lambda _{\beta }z_b\gg 1$ assuming that
$\lambda _{\beta } z_a\lesssim 1$. Using the asymptotic formulae for
the Bessel modified functions containing in the argument $z_b$, we
find the following result
\begin{eqnarray}
p^{(a)}_{{\mathrm{(int)}}\beta }&\approx &
\frac{k_D^{D+1}z_a^D}{2^{D_1+1} \pi ^{\frac{D_1}{2}-1}}
\frac{(\lambda _{\beta }/z_b)^{\frac{D_1}{2}} e^{-2\lambda _{\beta
}z_b}}{c_b(\lambda _{\beta }z_b)\bar K_{\nu }^{(a)
2}(\lambda _{\beta }z_a)} F_{\beta }^{(a)}(\lambda _{\beta }z_a), \label{pintlargezb1}\\
p^{(b)}_{{\mathrm{(int)}}\beta }&\approx &
\frac{(k_Dz_b)^{D+1}}{2^{D_1}\pi
^{\frac{D_1}{2}-1}z_b^{\frac{D_1}{2}}} \frac{\lambda _{\beta
}^{\frac{D_1}{2}+1}e^{-2\lambda _{\beta }z_b}}{(A_b+\lambda _{\beta
}z_bB_b)^2} \frac{\bar I_{\nu }^{(a)}(\lambda _{\beta }z_a)}{\bar
K_{\nu }^{(a)}(\lambda _{\beta }z_a)} F_{\beta }^{(b)} (\lambda
_{\beta }z_b) .\label{pintlargezb2}
\end{eqnarray}
This limit corresponds to the interbrane distances much larger
compared with the AdS curvature radius and inverse KK masses,
measured by an observer on the left brane: $b-a\gg 1/k_D, 1/\lambda
_{\beta }^{(a)}$. For a single parameter manifold $\Sigma $ with
length scale $L$ and $(b-a)\gg L_{a}$ these conditions are satisfied
for all nonzero KK modes.

In the limit $z_a\lambda _{\beta }\ll 1$ for fixed $z_b\lambda
_{\beta }$, by using the asymptotic formulae for the Bessel modified
functions for small values of the argument, one finds
\begin{eqnarray}
p^{(a)}_{{\mathrm{(int)}}\beta }&\approx &
\frac{k_D^{D+1}z_a^{D+2\nu }(A_a-\nu B_a)^{-2}F_{\beta
}^{(a)}(0)}{2^{D_1+2\nu -2} \pi ^{\frac{D_1}{2}}\Gamma \left(
\frac{D_1}{2}\right) \Gamma ^2(\nu )} \int _{\lambda _{\beta
}}^{\infty } du\, u^{2\nu +1}(u^2-\lambda _{\beta
}^2)^{\frac{D_1}{2}-1}
\frac{\bar K_{\nu }^{(b)}(uz_b)}{\bar I_{\nu }^{(b)}(uz_b)}, \label{pintsmallza1}\\
p^{(b)}_{{\mathrm{(int)}}\beta }&\approx & \frac{2^{1-D_1-2\nu
}k_D^{D+1}z_b^{D}z_a^{2\nu }}{\pi ^{\frac{D_1}{2}}\Gamma \left(
\frac{D_1}{2}\right) \nu \Gamma ^2(\nu ) c_a(\nu )} \int _{\lambda
_{\beta }}^{\infty } du\, u^{2\nu +1} (u^2-\lambda _{\beta
}^2)^{\frac{D_1}{2}-1}\frac{F_{\beta }^{(b)} (uz_b)}{\bar I_{\nu
}^{(b)2}(u z_b)} .\label{pintsmallza2}
\end{eqnarray}
In this case the KK masses measured by an observer on the brane at
$y=a$ are much less than the AdS energy scale, $\lambda _{\beta
}^{(a)}\ll k_D$, and the interbrane distance is much larger than the
AdS curvature radius. In particular, substituting $\lambda _{\beta
}=0$, from these formulae we obtain the asymptotic behavior for the
contribution of the zero mode to the interaction forces between the
branes in the limit $z_a/z_b\ll 1$. From formulae
(\ref{pintsmallza1}) and (\ref{pintsmallza2}) it follows that, in
dependence of values of the coefficients in Robin boundary
conditions, for large distances the interaction forces can be either
attractive or repulsive. In particular, for Dirichlet boundary
condition on the brane at $y=b$, one has
$p^{(a)}_{{\mathrm{(int)}}\beta }<0$ for $F_{\beta }^{(a)}(0)<0$ and
$p^{(b)}_{{\mathrm{(int)}}\beta }<0$ for $|A_a/B_a|>\nu $, and under
these conditions for $B_a\neq 0$ we have an interesting situation
when the interaction forces are repulsive for small distances (see
Eq. (\ref{pintsmalldist})) and are attractive for large distances.

From the formulae given above it follows that in the limit when the
right brane tends to the AdS horizon, $z_b\to \infty $, the force
$p^{(a)}_{{\mathrm{(int)}}\beta }$ vanishes as $e^{-2\lambda _{\beta
}z_b}/z_b^{D_1/2}$ for the nonzero KK mode along $\Sigma $ and as
$z_b^{-D_1-2\nu }$ for the zero mode. The effective pressure
$p^{(b)}_{{\mathrm{(int)}}\beta }$ vanishes as
$z_b^{D_2+D_1/2+1}e^{-2\lambda _{\beta }z_b}$ for the nonzero KK
mode and behaves as $z_b^{D_2-2\nu }$ for the zero mode. In the
limit when the left brane tends to the AdS boundary the contribution
of a given KK mode vanishes as $z_a^{D+2\nu }$ for the pressure
$p^{(a)}_{{\mathrm{(int)}}\beta }$ and as $z_a^{2\nu }$ for the
pressure $p^{(b)}_{{\mathrm{(int)}}\beta }$. For small values of the
AdS curvature radius corresponding to strong gravitational fields,
assuming $\lambda _{\beta }z_a \gg 1$ and $\lambda _{\beta
}(z_b-z_a) \gg 1$, we can estimate the contribution of the nonzero
KK modes to the vacuum interaction forces by formula
(\ref{pintlargeKK1}). In particular, for the case of a single
parameter internal space with the length scale $L$, under the
assumed conditions the length scale of the internal space measured
by an observer on the brane at $y=a$ is much smaller compared to the
AdS curvature radius, $L_{a}\ll k_{D}^{-1}$. If $L_{a}\gtrsim
k_{D}^{-1}$ one has $\lambda _{\beta }z_a\lesssim 1$ and to estimate
the contribution of the interference part we can use formulae
(\ref{pintlargezb1}) and (\ref{pintlargezb2}), and the suppression
is stronger compared with the previous case. For the zero KK mode,
under the condition $k_D(b-a)\gg 1$ we have $z_a/z_b\ll 1$ and to
the leading order the corresponding interaction forces are described
by relations (\ref{pintsmallza1}) and (\ref{pintsmallza2}). From
these formulae it follows that the interaction forces integrated
over the internal space behave as $k_{D}^{D_1+1}\exp [(D_1\delta
_{j}^{a}+2\nu )k_{D}(a-b)]$ for the brane at $y=j$ and are
exponentially suppressed. Note that in the model without the
internal space we have similar behavior with $\nu $ replaced by $\nu
_{1}$ and for a scalar field with $\zeta <\zeta _{D+D_1+1}$ the
suppression is relatively weaker.

\section{An example: $\Sigma =S^1$}

\label{sec:exampS1}

To make our discussion concrete, here we consider a simple example
with $\Sigma =S^1$. In this case the bulk corresponds to the
$AdS_{D+1}$ spacetime with one compactified dimension $X$. The
corresponding normalized eigenfunctions and eigenvalues are as
follows
\begin{equation}\label{psibetS1}
\psi _{\beta }(X)=\frac{1}{\sqrt{L}}e^{2\pi i \beta X/L},\quad
\lambda _{\beta }=\frac{2\pi }{L}|\beta |,\quad \beta =0,\pm 1,\pm
2, \ldots ,
\end{equation}
where $L$ is the length of the compactified dimension. The
boundary-free part of the vacuum energy-momentum tensor can be
evaluated on the base of formula (\ref{vevEMT1pl}) by using the
Wightman function and the VEV of the field square from
\cite{Saha05I}. The Wightman function is determined by the term in
formula (\ref{WF1bran}) with the first integral in the figure
braces. The application of the Abel-Plana formula to the series over
$\beta $ allows to present the boundary-free part of the
energy-momentum tensor in the form (no summation over $M$)
\begin{equation}\label{TMNS11}
\langle T_M^N\rangle ^{(0)}= \langle T_M^N\rangle _{AdS_{D+1}}^{(0)}
+ \delta _{M}^{N}\int_{0}^{\infty }du\,
\sum_{l=0}^{1}S_{(M)}^{(l)}f_{\nu }^{(l)}(u,z/L) ,
\end{equation}
where we use the following notations
\begin{eqnarray}
 S_{(\mu )}^{(0)}&=& \left( \frac{1}{4}-\zeta \right) ue^{u} \frac{(u-D-1)e^{u}
 +u+D+1}{(e^{u}-1)^2}-\frac{\zeta ue^{u}}{e^{u}-1} +D\zeta , \; S_{(\mu )}^{(1)}=
 -\frac{z^{2} u^{2}}{2L^{2}}, \\
 S_{(D-1)}^{(0)}&=& S_{(\mu )}^{(0)}+\frac{z^{2} u^{2}}{L^{2}} ,\; S_{(D-1)}^{(1)}=0,
  \; S_{(D)}^{(1)}=
 -D_1S_{(\mu )}^{(1)}, \\
 S_{(D)}^{(0)}&=& -\frac{u}{4}e^{u} \frac{(u-D-1)e^{u}
 +u+D+1}{(e^{u}-1)^2}+\frac{D\zeta ue^{u}}{e^{u}-1} -D^2\zeta +\frac{m^2}{k_D^2}-
 \frac{z^{2} u^{2}}{L^{2}} ,
\end{eqnarray}
and
\begin{equation}\label{fnul}
f_{\nu }^{(l)}(u,z/L)=\frac{2k_D^{D+1}}{\pi ^{\frac{D-1}{2}}}
\frac{{}_{1}F_{2}\left( \nu +
  \frac{1}{2}; \nu +l+ \frac{D+1}{2}, 2\nu +1;
  -\frac{z^2u^2}{L^2}\right) }{\Gamma
(\nu +1)\Gamma \left( \nu +l+\frac{D+1}{2}\right)u(e^u-1)} \left(
\frac{uz}{2L}\right) ^{D+2\nu } ,
\end{equation}
with the hypergeometric function ${}_1F_{2}$. The first term on the
right of formula (\ref{TMNS11}) is the corresponding tensor in
$AdS_{D+1}$ bulk without boundaries and the second term is induced
by the compactness of $X$ direction. The latter is finite and the
renormalization procedure is needed for the term $\langle
T_M^N\rangle _{AdS_{D+1}}^{(0)}$ only. The renormalized value of
this tensor does not depend on the spacetime point and is
well-investigated in literature. For this reason below we will be
concentrated on the second term. For $z\ll L$ to the leading order
one has
\begin{equation}\label{TMNS1smallz}
\langle T_M^N\rangle ^{(0)}\approx \langle T_M^N\rangle
_{AdS_{D+1}}^{(0)} -2 \delta _{M}^{N} \frac{k_D^{D+1}\zeta
_{D}^{(+)}}{\pi ^{\frac{D}{2}}\Gamma (\nu )}\zeta _{R}(D+2\nu
)\Gamma \left( \frac{D}{2}+\nu \right) \left( \frac{z}{L}\right)
^{D+2\nu }  ,
\end{equation}
for $M=0,1,\ldots ,D-1$ and $\zeta _{R}(z)$ being the Riemann zeta
function. The corresponding formula for the component $\langle
T_{D}^{D}\rangle ^{(0)}$ differs from (\ref{TMNS1smallz}) by an
additional coefficient $-D/2\nu $ in the second term on the right of
this formula. This directly follows from the continuity equation
(\ref{conteq1}). Hence, the second term on the right of formula
(\ref{TMNS11}) vanishes at the AdS boundary as $z^{D+2\nu }$. For
both minimally and conformally coupled cases the energy density
corresponding to this term is positive. Note that in this limit the
vacuum stresses along compactified and uncompactified directions are
isotropic. Of course, we could expect the vanishing of the second
term on the right hand side of Eq. (\ref{TMNS11}) when $z/L\to 0$,
as this corresponds to the decompactification limit for the internal
space. For $z\gg L$, by using the asymptotic formula for the
hypergeometric function for large values of the argument, to the
leading order one finds
\begin{equation}\label{TMNS1largez}
\langle T_M ^N \rangle ^{(0)}\approx \langle T_M ^N \rangle
_{AdS_{D+1}}^{(0)}-\delta_{M}^{N}\frac{\zeta _{R}(D+1)}{\pi
^{\frac{D+1}{2}}}\Gamma \left( \frac{D+1}{2}\right) \left(
\frac{k_Dz}{L}\right) ^{D+1} ,
\end{equation}
for $M\neq D-1$. The corresponding formula for
${}_{D-1}^{D-1}$--component differs from (\ref{TMNS1largez}) by an
additional coefficient $-D$ in the second term on the right. In
particular, from here it follows that the energy density
corresponding to the second term on the right of Eq. (\ref{TMNS11})
is negative near the AdS horizon and diverges as $z^{D+1}$. The same
is the case for $\langle T_0^0\rangle ^{(0)}$ as the energy density
corresponding to the first term does not depend on $z$ and the total
energy density is dominated by the second term. The limit $z\gg L$
is realized, in particular, when $k_D\to 0$ for fixed $y$ and $L$.
In this case the first term on the right of Eq. (\ref{TMNS1largez})
vanishes and from this formula we obtain the standard result for the
Casimir energy-momentum tensor in $R^{(D-1,1)}\times S^{1}$.
Combining the asymptotic formulae (\ref{TMNS1smallz}) and
(\ref{TMNS1largez}), we see that for a scalar field with $\zeta
<\zeta ^{(+)}_{D}$ (in particular, this is the case for minimally
and conformally coupled scalars) the energy density corresponding to
the second term on the right of formula (\ref{TMNS11}) tends to zero
for small values of the ratio $z/L$ being positive and tends to
$-\infty $ for large values of this ratio. Hence, it has a maximum
at some intermediate value of $z/L$. As the first term on the right
of Eq. (\ref{TMNS11}) is constant, the same is true for the
boundary-free total energy density.

\begin{figure}[tbph]
\begin{center}
\begin{tabular}{cc}
\epsfig{figure=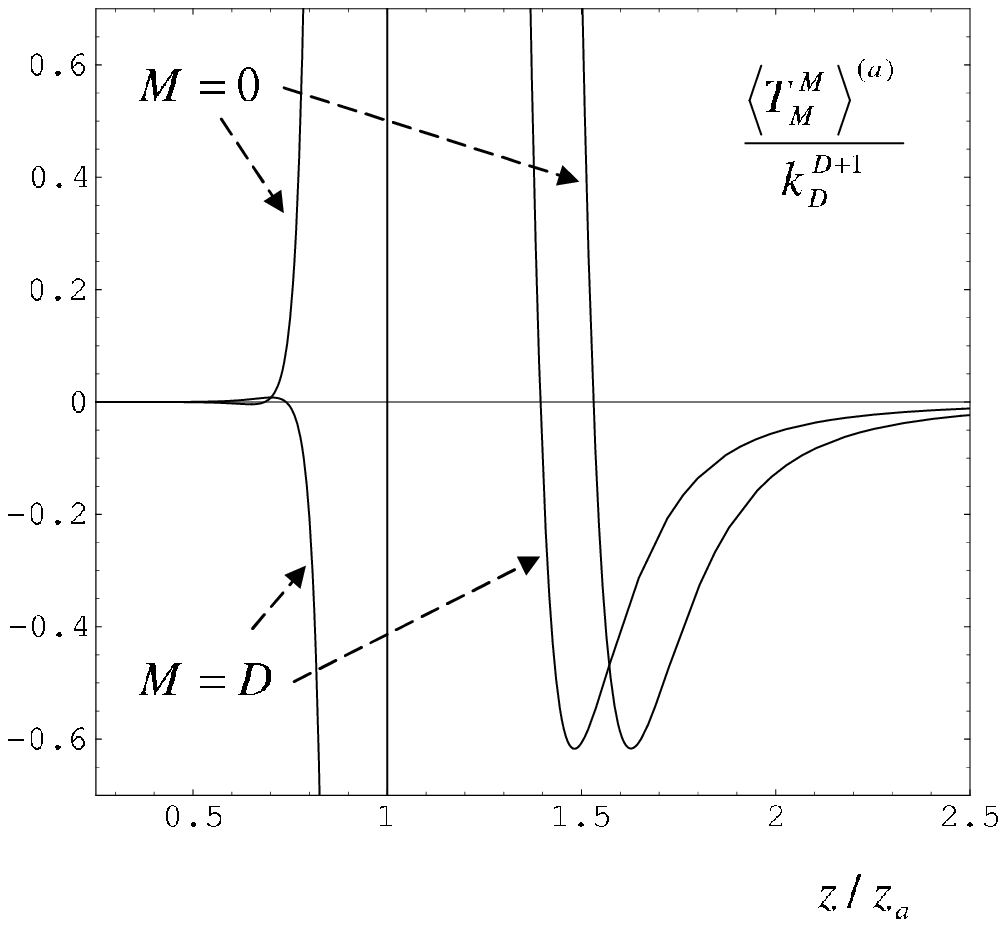,width=6.5cm,height=6cm}& \quad
\epsfig{figure=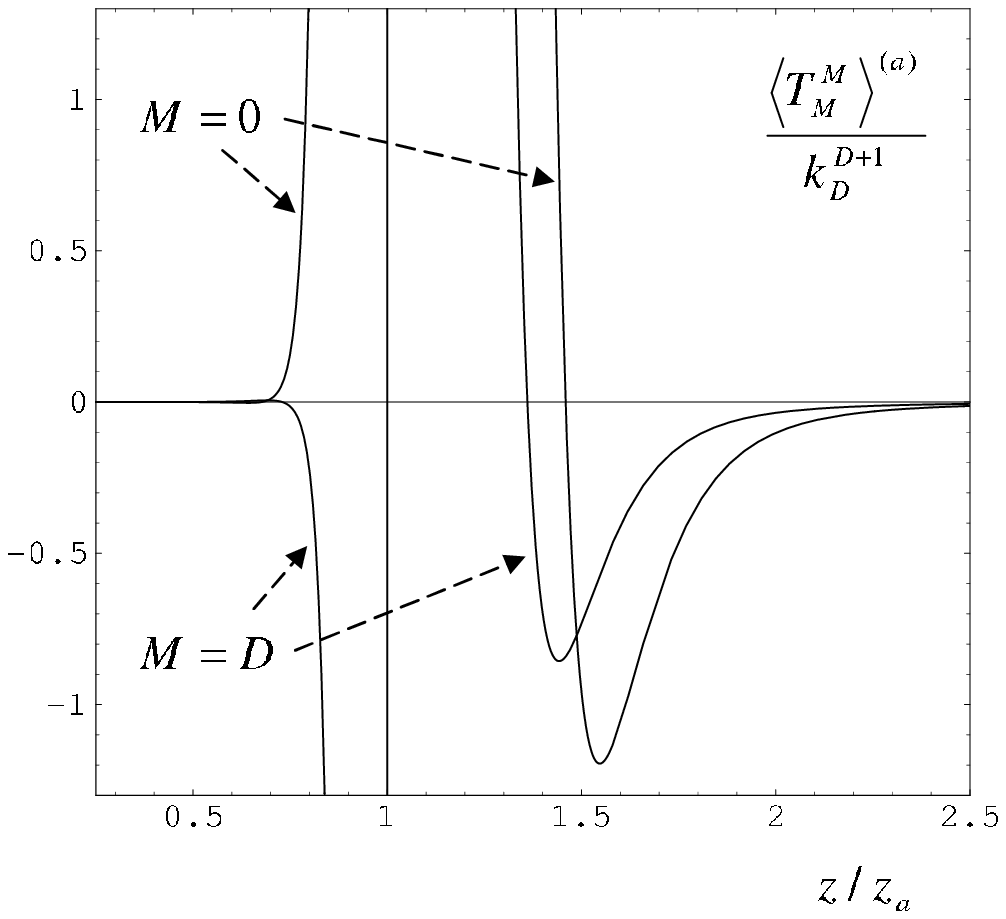,width=6.5cm,height=6cm}
\end{tabular}
\end{center}
\caption{Brane induced vacuum energy density $\langle T^0_0\rangle
^{(a)}/k_{D}^{D+1}$ and stress $\langle T^D_D\rangle
^{(a)}/k_{D}^{D+1}$ in units of $k_D^{D+1}$ as functions on $z/z_a$
for a minimally coupled massless scalar in $D=5$. The left panel is
for $L/z_a=1$ and the right one is for $L/z_a=2$.} \label{fig1}
\end{figure}

For the internal space $S^{1}$ the brane induced VEV of the
energy-momentum tensor and the vacuum interaction forces between the
branes are obtained from general formulae given in previous sections
by the replacements
\begin{equation}\label{replS1}
\sum_{\beta }|\psi _{\beta }(X)|^2\to
\frac{2}{L}\sideset{}{'}{\sum}_{\beta =0}^{\infty }, \quad \lambda
_{\beta }\to \frac{2\pi }{L}|\beta |, \quad D_1\to D-1 ,
\end{equation}
where the prime means that the summand $\beta =0$ should be taken
with the weight 1/2. From the general analysis given above it
follows that under the conditions $z\gg L,z_a$, the main
contribution into the part of the vacuum energy-momentum tensor
induced by a single brane comes from the zero mode and this part
behaves as $z^{1-2\nu }$. Comparing with (\ref{TMNS1largez}), we see
that for the geometry of a single brane, near the AdS horizon the
total vacuum energy-momentum tensor is dominated by the second term
on the right of Eq. (\ref{TMNS11}). Near the AdS boundary, $z\ll
L,z_a$, the single brane induced term in the vacuum energy-momentum
tensor behaves like $z^{D+2\nu }$ (see Eq. (\ref{TMNsplsmallz})). By
taking into account the estimate (\ref{TMNS1smallz}), we see that in
this case the vacuum energy-momentum tensor is dominated by the
first term on the right of (\ref{TMNS11}). For the points near the
brane the main contribution into the energy-momentum tensor comes
from the brane induced term. As an illustartion, in figure
\ref{fig1} we have plotted single brane induced VEVs for the vacuum
energy density and ${}^{D}_{D}$--stress as functions on $z/z_a$ for
the internal space with $L/z_a=1$ (left panel) and $L/z_a=2$ (right
panel) in the case of $D=5$ minimally coupled massless scalar field
with the ratio of Robin coefficients $B_a/\tilde A_a=0.15$. In this
case for large values $z/z_a$ the brane induced part of the
energy-momentum tensor behaves as $(z/z_{a})^{-4}$. For small values
of this ratio one has the behavior $(z/z_{a})^{10}$.

\begin{figure}[tbph]
\begin{center}
\begin{tabular}{cc}
\epsfig{figure=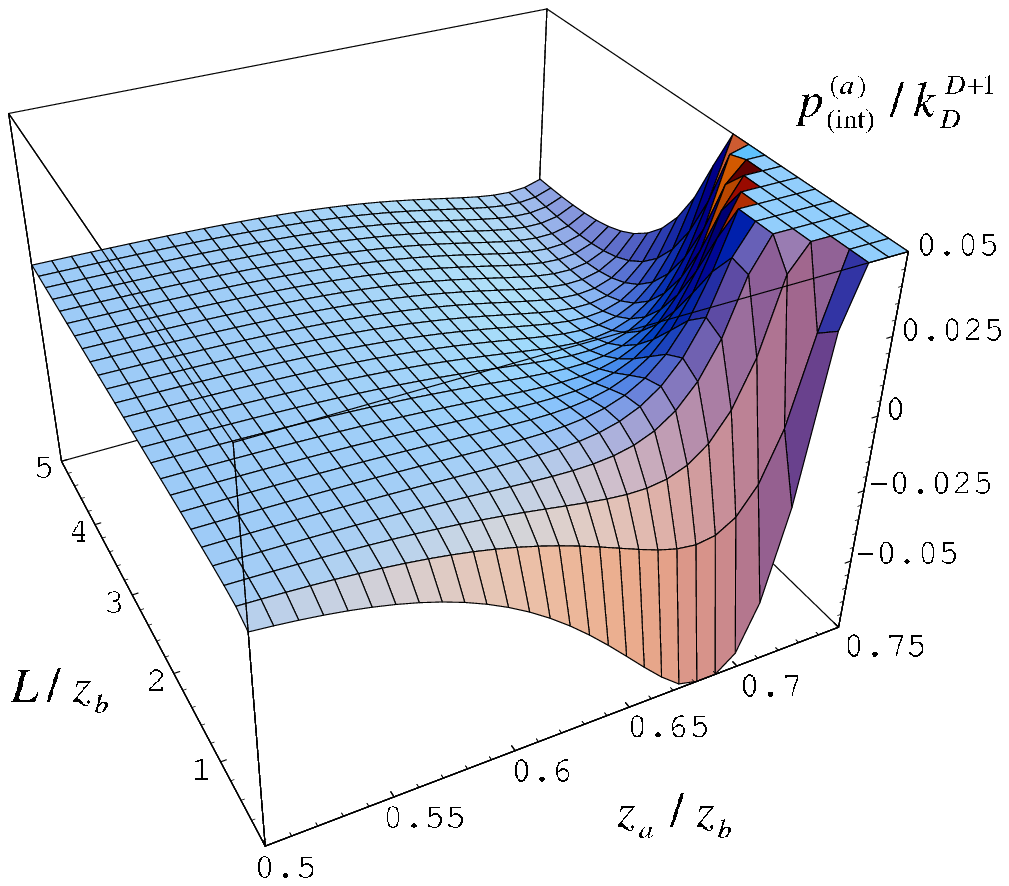,width=6.5cm,height=6.5cm}& \quad
\epsfig{figure=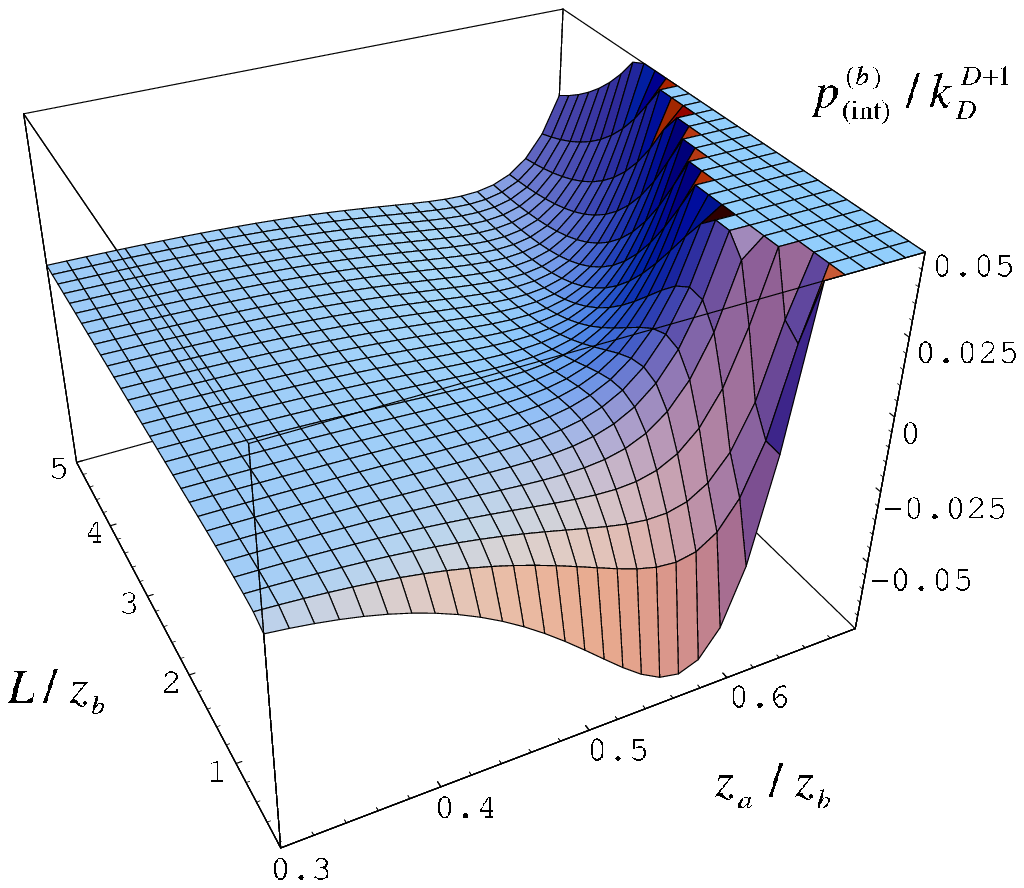,width=6.5cm,height=6.5cm}
\end{tabular}
\end{center}
\caption{The $D=5$ vacuum effective pressures
$p^{(j)}_{{\mathrm{int}}}/k_D^{D+1}$, $j=a,b$, in units of
$k_D^{D+1}$ for a massless minimally coupled scalar field as
functions on $L/z_b$ and $z_a/z_b$. The values for the Robin
coefficients are $B_a=0$ and $B_b/\tilde A_b=0.15$. Left panel
corresponds to $j=a$ and right panel corresponds to $j=b$.}
\label{fig2}
\end{figure}

In figure \ref{fig2} we present the vacuum interaction forces in the
geometry of two branes as functions on the size of the internal
space and interbrane distance for a $D=5$ minimally coupled massless
scalar field with the Robin coefficients $B_a=0$ and $B_b/\tilde
A_{b} =0.15$. In this example the effective pressures
$p^{(j)}_{{\mathrm{(int)}}}$ are positive for small interbrane
distances and are negative for large distances leading to the
repulsive and attractive interaction forces respectively.

From the point of view of embedding the Randall-Sundrum model into
the string theory and also in the discussions of the holographic
principle, the case of the internal space $\Sigma =S^{D_2}$ is of
special interest. The corresponding eigenfunctions $\psi _{\beta
}(X)$ are expressed in terms of spherical harmonics of degree $l$,
$l=0,1,2,\ldots $. For the internal space with radius $R_0$ the VEVs
of the energy-momentum tensor and the vacuum interaction forces are
obtained from the general formulae in previous sections by the
replacements
\begin{eqnarray}
\sum_{\beta }|\psi _{\beta }(X)|^2 &\to & \frac{\Gamma \left(
\frac{D_2+1}{2}\right)}{2 \pi
^{\frac{D_2+1}{2}}R_0^{D_2}}\sum_{l=0}^{\infty }
(2l+D_2-1)\frac{\Gamma (l+D_2-
1)}{l! \Gamma (D_2)}, \label{SD1psi} \\
  \lambda _{\beta } &\to & \frac{1}{R_0}\sqrt{l(l+D_2-1)+\zeta
  D_2(D_2-1)} \label{SD1lambda} ,
\end{eqnarray}
with the factor under the summation sign on the right in Eq.
(\ref{SD1psi}) being the degeneracy of the angular mode with a given
$l$.

\section{Conclusion}

\label{sec:Conc}

In this paper we continue the investigation of local quantum effects
in higher dimensional brane models on the bulk of topology
$AdS_{D_1+1}\times \Sigma $ with a warped internal subspace $\Sigma
$ and the line element (\ref{metric}). The case of bulk scalar field
with general curvature coupling parameter and satisfying Robin
boundary conditions on two codimension one branes is considered. In
section \ref{sec:EMT1pl} the model with a single brane is discussed.
By using the Wightman function and the VEV of the field square from
Ref. \cite{Saha05I}, in the general case of the extra space we
derived formulae for the VEV of the energy-momentum tensor in both
regions on the right and on the left from the brane, given by
expressions (\ref{EMT1bounda}) and (\ref{Tik1plnewleft})
respectively. Unlike to the case of purely AdS bulk, here the VEVs
in addition to the distance from the brane depend also on the
position of the brane in the bulk. In the limit when the AdS
curvature radius tends to infinity we derive the formula for the
vacuum energy-momentum tensor for parallel plates on the background
spacetime with topology $R^{(D_1,1)}\times \Sigma $. In this limit
for a homogeneous internal space ${}^{D}_{D}$--component of the
brane induced part in the VEV of the energy-momentum tensor
vanishes. Further we have investigated various limiting cases when
the expression for the brane induced VEV is simplified. For the
points on the brane the vacuum energy-momentum tensor diverges. The
leading term in the corresponding asymptotic expansion near the
brane is given by formula (\ref{TMNsplnear}). Near the brane the
total vacuum energy-momentum tensor is dominated by the brane
induced part and has opposite signs for Dirichlet and non-Dirichlet
boundary conditions. Near the brane ${}^{D}_{D}$-- and
${}^{i}_{D}$--components of this tensor have opposite signs in the
regions $y<a$ and $y>a$. For $\zeta >\zeta _{D_1}$ ($\zeta <\zeta
_{D_1}$) the energy density is positive (negative) for Dirichlet
boundary condition and is negative (positive) for non-Dirichlet
boundary condition. For large distances from the brane in the region
$z>z_a $ the contribution of a given mode along $\Sigma $ with
nonzero KK mass is exponentially suppressed by the factor
$e^{-2\lambda _{\beta }z}$. For the zero mode the brane induced VEV
near the AdS horizon behaves as $z^{D_2-2\nu }$. In the purely AdS
bulk ($D_2=0$) this VEV vanishes on the horizon for $\nu >0$. For an
internal spaces with $D_2>2\nu $ the VEV diverges on the horizon.
The VEV integrated over the internal space (see Eq.
(\ref{TMNintegrated})) vanishes on the AdS horizon for all values
$D_2$ due to the additional warp factor coming from the volume
element. For the points near the AdS boundary, the brane induced VEV
vanishes as $z^{D+2\nu }$ for diagonal components and as $z^{D+2\nu
+2}$ for the ${}_D^i$--component. For small values of the length
scale for the internal space, the contribution of nonzero KK masses
is exponentially suppressed and the main contribution into the brane
induced energy-momentum tensor comes from the zero mode. In the
opposite limit, when the length scale of the internal space is
large, to the leading order the vacuum energy-momentum tensor
reduces to the corresponding result for a brane in the bulk
$AdS_{D+1}$ given in Ref. \cite{Saha04a}. For strong gravitational
fields corresponding to small values of the AdS curvature radius,
the contribution from nonzero KK modes along $\Sigma $ is
exponentially suppressed by the factor $e^{-2\lambda _{\beta
}|z-z_a|}$. For the zero KK mode the components of the brane induced
vacuum energy-momentum tensor behave like
$k_{D}^{D_1+1}e^{D_2k_Dy}\exp [(D_1+2\nu )k_D(y-a)]$ in the region
$y<a$ and like $k_{D}^{D_1+1}e^{D_2k_Dy}\exp [2\nu k_D(a-y)]$ in the
region $y>a$. The corresponding quantities integrated over the
internal space contain additional factor $e^{-D_2k_Dy}$ coming from
the volume element and are exponentially small in both regions. For
fixed values of the other parameters, the brane induced VEV in the
region $z>z_a$ vanishes as $z_a^{2 \nu }$ when the brane position
tends to the AdS boundary. When the brane position tends to the AdS
horizon, $z_a\to \infty $, for massive KK modes along $\Sigma $ the
VEV of the energy-momentum tensor in the region $z<z_a$ is
suppressed by the factor $e^{-2z_a\lambda _{\beta }}$. For the zero
mode in the same limit the suppression is power-law with respect to
$z_a$.

The geometry of two branes we consider in section \ref{sec:EMT2pl}.
The VEVs in the region between the branes is presented in the form
(\ref{TMNtwopl1}) with separated boundary-free, single branes and
interference parts. The latter is given by formula (\ref{intpart})
and is finite everywhere including the points on the branes. The
surface divergences are contained in the single brane parts only. We
have explicitly checked that the both single brane and interference
parts separately satisfy the continuity equation and are traceless
for a conformally coupled massless scalar. The possible trace
anomalies are contained in the boundary-free parts. In the limit
$k_D\to 0$ we derive the corresponding results for two parallel
Robin plates in the bulk $R^{(D_1,1)}\times \Sigma $. For small
values of the length scale of the internal space corresponding to
large KK masses, the interference part in the VEV of the
energy-momentum tensor is estimated by formula
(\ref{TMN2pllargelamb1}) and is exponentially suppressed. For large
interbrane distances compared with the AdS curvature radius
$k_D^{-1}$ we have approximate formula (\ref{TMN2pllargezbn}) for
the nonzero KK modes along $\Sigma $. This limit is realized in
braneworld scenarios for the solution of the hierarchy problem. The
interference part vanishes as $z_a^{2\nu }$ when the left brane
tends to the AdS boundary. Under the condition $z\ll z_b$ an
additional suppression factor appears in the form $(z/z_b)^{D_1}$
for ${}^{D}_{D}$--component and in the form $(z/z_b)^{D_1+2\alpha
_1}$ for the other components, where $\alpha _{1}=\min (1,\nu )$. In
a higher dimensional generalization of the Randall-Sundrum
braneworld based on the bulk $AdS_{D_1+1}\times \Sigma $ with
orbifolded $y$-direction the VEV of the energy-momentum tensor is
obtained from the formulae in this paper with an additional factor
1/2. For untwisted scalar field the corresponding Robin coefficients
are related to the surface mass parameters by formulae
(\ref{AtildeRS}). For twisted scalar Dirichlet boundary conditions
are obtained. Note that in the present paper we consider the VEV of
the bulk energy-momentum tensor. On manifolds with boundaries in
addition to this part, the energy-momentum tensor contains a
contribution located on on the boundary (for the expression of the
surface energy-momentum tensor in the case of arbitrary bulk and
boundary geometries see Ref. \cite{Saha04c}). As it has been
discussed in Refs.
\cite{Rome02,Saha04c,Kenn80,Saha01,Rome01,Full03}, the surface part
of the energy-momentum tensor is essential in considerations of the
relation between local and global characteristics in the Casimir
effect. The vacuum expectation value of the surface energy-momentum
tensor for the geometry of two parallel branes in $AdS_{D+1}$ bulk
is evaluated in Ref. \cite{Saha04d}. In particular, it has been
shown that for large distances between the branes the induced
surface densities give rise to an exponentially suppressed
cosmological constant on the brane. The investigation of the surface
densities and induced cosmological constant on the branes for the
bulk $AdS_{D_1+1}\times \Sigma $ will be reported in the forthcoming
paper.

The vacuum effective pressure acting on the branes is determined by
the ${^{D}_{D}}$--component of the energy-momentum tensor and can be
separated into single brane and second brane induced parts. The
first one is divergent and needs additional renormalization. The
corresponding procedure lies on the same line as the evaluation of
the surface energy-momentum tensor and will be discussed in the
forthcoming paper. Here we concentrate on the second brane induced
part which is finite for all nonzero interbrane distances and is not
changed by the regularization and renormalization procedure. For the
brane at $z=z_j$ this term is determined by formula (\ref{pintj}).
For Dirichlet scalar the corresponding vacuum forces are attractive
for all interbrane distances. Taking the limit $k_D\to 0$ we
obtained the result for the bulk $R^{(D_1-1,1)}\times \Sigma $,
given by the right hand side of Eq. (\ref{pintkD0}). In this case,
for a homogeneous internal space the interaction forces are the same
even in the case of different Robin coefficients for separate
branes. For the modes along $\Sigma $ with large KK masses, the
interaction forces are exponentially small. In particular, for
sufficiently small length scale of the internal space this is the
case for all nonzero KK modes and the main contribution to the
interaction forces comes from the zero mode. For small interbrane
distances, to the leading order the interaction forces are given by
formula (\ref{pintsmalldist}). In this limit they are repulsive for
Dirichlet boundary condition on one brane and non-Dirichlet boundary
condition on the another and are attractive for other cases. For
small interbrane distances the contribution of the interaction term
dominates the single brane parts, and the same is the case for the
total vacuum forces acting on the branes. When the right brane tends
to the AdS horizon, $z_b\to \infty $, the interaction force acting
on the left brane vanishes as $e^{-2\lambda _{\beta
}z_b}/z_b^{D_1/2}$ for the nonzero KK mode and like $z_b^{-D_1-2\nu
}$ for the zero mode. In the same limit the corresponding force
acting on the right brane behaves as $z_b^{D_2+D_1/2+1}e^{-2\lambda
_{\beta }z_b}$ for the nonzero KK mode and like $z_b^{D_2-2\nu }$
for the zero mode. In the limit when the left brane tends to the AdS
boundary the contribution of a given KK mode into the vacuum
interaction force vanishes as $z_a^{D+2\nu }$ and as $z_a^{2\nu }$
for the left and right branes, respectively. For small values of the
AdS curvature radius corresponding to strong gravitational fields,
for nonzero KK modes under the conditions $\lambda _{\beta }z_a \gg
1$ and $\lambda _{\beta }(z_b-z_a) \gg 1$, the contribution to the
interaction forces is suppressed by the factor $e^{-2\lambda _{\beta
}(z_b-z_a)}$. For the zero KK mode, the corresponding interaction
forces integrated over the internal space behave as
$k_{D}^{D_1+1}\exp [(D_1\delta _{j}^{a}+2\nu )k_{D}(a-b)]$ for the
brane at $y=j$ and are exponentially small. In the model without the
internal space we have similar behavior with $\nu $ replaced by $\nu
_{1}$ and for a scalar field with $\zeta <\zeta _{D+D_1+1}$ the
suppression is relatively weaker.

As an example of application of the general results, in section
\ref{sec:exampS1} we consider a special case of the internal space
with $\Sigma =S^{1}$. By using the Abel-Plana summation formula, the
boundary-free part of the vacuum energy-momentum tensor is presented
in the form (\ref{TMNS11}), where the first term on the right is the
corresponding tensor in $AdS_{D+1}$ bulk without boundaries. The
second term is induced by the compactness of the $X$ direction and
behaves as $z^{D+2\nu }$ for $z\ll L$ and like $z^{D+1}$ for $z\gg
L$. For minimally and conformally coupled scalar the corresponding
energy density is positive in the first case and negative in the
second case and has a maximum at some intermediate value for $z/L$.
The total energy-momentum tensor is dominated by the first term on
the right of Eq. (\ref{TMNS11}) near the AdS boundary and by the
second term near the horizon. For the points near the branes the
brane induced parts dominate. In the case of internal space $\Sigma
=S^1$ the latter are obtained from the general formulae by
replacements (\ref{replS1}). We also comment on the generalization
to the case of the internal space $\Sigma =S^{D_2}$.

\section*{Acknowledgments}

The work was supported by the CNR-NATO Senior Fellowship, by ANSEF
Grant No. 05-PS-hepth-89-70, and in part by the Armenian Ministry of
Education and Science, Grant No. 0124. The author acknowledges the
hospitality of the Abdus Salam International Centre for Theoretical
Physics (Trieste, Italy) and Professor Seif Randjbar-Daemi for his
kind support.

\end{document}